\documentclass[sn-mathphys-num]{sn-jnl}

\usepackage{graphicx}%
\usepackage{multirow}%
\usepackage{amsmath,amssymb,amsfonts}%
\usepackage{amsthm}%
\usepackage{mathrsfs}%
\usepackage[title]{appendix}%
\usepackage{xcolor}%
\usepackage{textcomp}%
\usepackage{manyfoot}%
\usepackage{booktabs}%
\usepackage{algorithm}%
\usepackage{algorithmicx}%
\usepackage{algpseudocode}%
\usepackage{listings}%

\UseRawInputEncoding

\theoremstyle{thmstyleone}%
\theoremstyle{thmstyletwo}%

\theoremstyle{thmstylethree}%

\graphicspath{{./}{figures/}}

\begin{document}

\title[Article Title]{Continuous helium absorption from the leading and trailing tails of WASP-107\,b}


\author*[1,2]{\fnm{Vigneshwaran} \sur{Krishnamurthy}}\email{vigneshwaran.krishnamurthy@mcgill.ca}

\author[3]{\fnm{Yann} \sur{Carteret}}

\author[4,7]{\fnm{Caroline} \sur{Piaulet-Ghorayeb}}

\author[1,5]{\fnm{Jared} \sur{Splinter}}

\author[1,2]{\fnm{Dhvani} \sur{Doshi}}

\author[6,7]{\fnm{Michael} \sur{Radica}}

\author[6]{\fnm{Louis-Philippe} \sur{Coulombe}}

\author[6]{\fnm{Romain} \sur{Allart}}

\author[3]{\fnm{Vincent} \sur{Bourrier}}

\author[1,2,4]{\fnm{Nicolas} \sur{B. Cowan}}

\author[6]{\fnm{David} \sur{Lafreni\`ere}}

\author[6]{\fnm{Lo\"{i}c} \sur{Albert}}

\author[4]{\fnm{Lisa} \sur{Dang}}

\author[8]{\fnm{Ray} \sur{Jayawardhana}}

\author[9,10]{\fnm{Doug} \sur{Johnstone}}

\author[11]{\fnm{Lisa} \sur{Kaltenegger}}

\author[8,12]{\fnm{Adam B.} \sur{Langeveld}}

\author[3,6]{\fnm{Stefan} \sur{Pelletier}}

\author[13]{\fnm{Jason F.} \sur{Rowe}}

\author[4]{\fnm{Pierre-Alexis} \sur{Roy}}

\author[14]{\fnm{Jake} \sur{Taylor}}

\author[12]{\fnm{Jake D.} \sur{Turner}}


\affil[1]{\orgdiv{Trottier Space Institute}, \orgname{McGill University}, \orgaddress{\street{3550 rue University}, \city{Montr\'eal}, \postcode{H3A 2A7}, \state{QC}, \country{Canada}}}

\affil[2]{\orgdiv{Department of Physics}, \orgname{McGill University}, \orgaddress{\street{3600 rue University}, \city{Montr\'eal}, \postcode{H3A 2T8}, \state{QC}, \country{Canada}}}

\affil[3]{\orgdiv{Observatoire Astronomique}, \orgname{l’Universit\'e de Gen\`eve}, \orgaddress{\street{Chemin Pegasi 51b}, \city{Versoix}, \state{Geneva}, \country{Switzerland}}}

\affil[4]{\orgdiv{Institut Trottier de Recherche sur les Exoplanètes and Département de Physique}, \orgname{Université de Montréal}, \orgaddress{\street{1375 Avenue Thérèse-Lavoie-Roux}, \city{Montréal}, \postcode{H2V 0B3}, \state{QC}, \country{Canada}}}

\affil[5]{\orgdiv{Department of Earth \& Planetary Sciences}, \orgname{McGill University}, \orgaddress{\street{3450 rue University}, \city{Montr\'eal}, \postcode{H3A 0E8}, \state{QC}, \country{Canada}}}

\affil[6]{\orgdiv{Institut Trottier de Recherche sur les Exoplanètes}, \orgname{Université de Montréal}, \orgaddress{\street{1375 Avenue Thérèse-Lavoie-Roux}, \city{Montréal}, \postcode{H2V 0B3}, \state{QC}, \country{Canada}}}

\affil[7]{\orgdiv{Department of Astronomy \& Astrophysics}, \orgname{University of Chicago}, \orgaddress{\street{5640 South Ellis Avenue}, \city{Chicago}, \postcode{60637}, \state{IL}, \country{USA}}}

\affil[8]{\orgdiv{Department of Physics \& Astronomy}, \orgname{Johns Hopkins University}, \orgaddress{\street{3400 N. Charles Street}, \city{Baltimore}, \postcode{21218}, \state{MD}, \country{USA}}}

\affil[9]{\orgname{NRC Herzberg Astronomy and Astrophysics}, \orgaddress{\street{5071 West Saanich Rd}, \city{Victoria}, \postcode{V9E 2E7}, \state{BC}, \country{Canada}}}

\affil[10]{\orgdiv{Department of Physics \& Astronomy}, \orgname{University of Victoria}, \orgaddress{\city{Victoria}, \postcode{V8P 5C2}, \state{BC}, \country{Canada}}}

\affil[11]{\orgdiv{Carl Sagan Institute}, \orgname{Cornell University}, \orgaddress{\street{302 Space Science Building}, \city{Ithaca}, \postcode{14850}, \state{NY}, \country{USA}}}

\affil[12]{\orgdiv{Department of Astronomy and Carl Sagan Institute}, \orgname{Cornell University}, \orgaddress{\city{Ithaca}, \postcode{14850}, \state{NY}, \country{USA}}}

\affil[13]{\orgdiv{Department of Physics and Astronomy}, \orgname{Bishops University}, \orgaddress{\street{2600 Rue College}, \city{Sherbrooke}, \postcode{J1M 1Z7}, \state{QC}, \country{Canada}}}

\affil[14]{\orgdiv{Department of Physics}, \orgname{University of Oxford}, \orgaddress{\street{Parks Rd}, \city{Oxford}, \postcode{OX1 3PU}, \country{UK}}}


\abstract{The detection of helium escaping the atmosphere of exoplanets has revolutionized our understanding of atmospheric escape and exoplanetary evolution. Using high-precision spectroscopic observations from the \textit{James Webb Space Telescope} (\textit{JWST}) NIRISS-SOSS mode, we report the detection of significant helium absorption during the pre-transit phase of WASP-107\,b (17\,$\sigma$), as well as in the transit and post-transit phases. This unique continuous helium absorption begins approximately 1.5 hours before the planet's ingress and reveals the presence of an extended thermosphere. The observations show a maximum transit depth of 2.395\,\% $\pm$ 0.01\,\% near the helium triplet (36\,$\sigma$; at NIRISS-SOSS resolution $\sim$\,700). 
Our ellipsoidal model of the planetary thermosphere matches well the measured light curve suggesting an outflow extending to tens of planetary radii. Furthermore, we confidently detect water absorption ($\log_{10}$\,H$_2$O=$-2.5 \pm 0.6$), superimposed with a short-wavelength slope which we attribute to a prominent signature from unocculted stellar spots (5.2\,$\sigma$), rather than a small-particle haze slope. We place an upper limit on the abundance of K ($\log_{10}$\,K\,$<$\,-4.86, or K/H$<75 \times$ stellar) at 2\,$\sigma$, which is consistent with the O/H super-solar metallicity estimate. This study underscores the transformative potential of \textit{JWST} for tracing atmospheric and mass-loss processes, while offering a benchmark for future studies targeting helium escape and its implications for planetary evolution.}




\maketitle

\section{Main} \label{sec: main}

WASP-107\,b \citep{wasp107b_anderson2017} is an ultra-low-density super-Neptune (0.96 $\pm$ 0.03\,$\mathrm{R_{J}}$; 30.5 $\pm$ 1.7\,$M_{\oplus}$) orbiting a K6 star in a 5.72-day orbit. It is the first planet for which the helium in the extended atmosphere was detected using \textit{HST} \citep{spake2018}. Ground-based observations have revealed a comet-like tail structure in the evaporating outflow, evidenced by post-transit excess absorption of helium \citep{allart_high-resolution_2019, kirk2020_helium, spake_posttransit_2021}. Recent \textit{JWST} observations have shown that the planet has high internal heating ($>$\,345\,K), vigorous mixing (log$_\mathrm{10}$\,K$_\mathrm{zz}$ = 8.5--11.6 cm$^2$\,s$^{-1}$) and super-solar metallicity (10--40$\times$\,solar) \citep{wellbanks2024_nircam, sing2024_nirspec}.

Here, we present the first-ever observation of continuous pre- and post-transit excess helium absorption for any exoplanet, achieved with the \textit{JWST} Near Infrared Imager and Slitless Spectrograph \citep[NIRISS;][]{doyon2023_nirisspaper} in Single Object Slitless Spectroscopy mode \citep[SOSS;][]{albert2023_sosspaper}. The transit observation was conducted as part of GTO 1201 (PI: Lafreni{\`e}re) on June 11, 2023. The SOSS mode in NIRISS is specifically tailored for time-series observations (TSOs) such as exoplanet transit observations \citep{albert2023_sosspaper}. Our TSO spanned 6.2 hours, encompassing 2.33 hours of pre-transit baseline, the 2.75-hour transit, along with 1.12 hours of post-transit baseline. Each integration lasted 33 seconds, resulting in a total of 678 integrations. This observation of WASP-107\,b with NIRISS completes the 0.6-12\,$\mu$m spectral range, making WASP-107\,b only the second planet for which we have transit spectra with all \textit{JWST} instruments \citep{wellbanks2024_nircam, sing2024_nirspec} \citep[after WASP-39\,b;][]{carter2024_wasp39b_panch1, powell2024_wasp39b_miri}.

\begin{figure*}[]
  \centering
  \includegraphics[width=\linewidth]{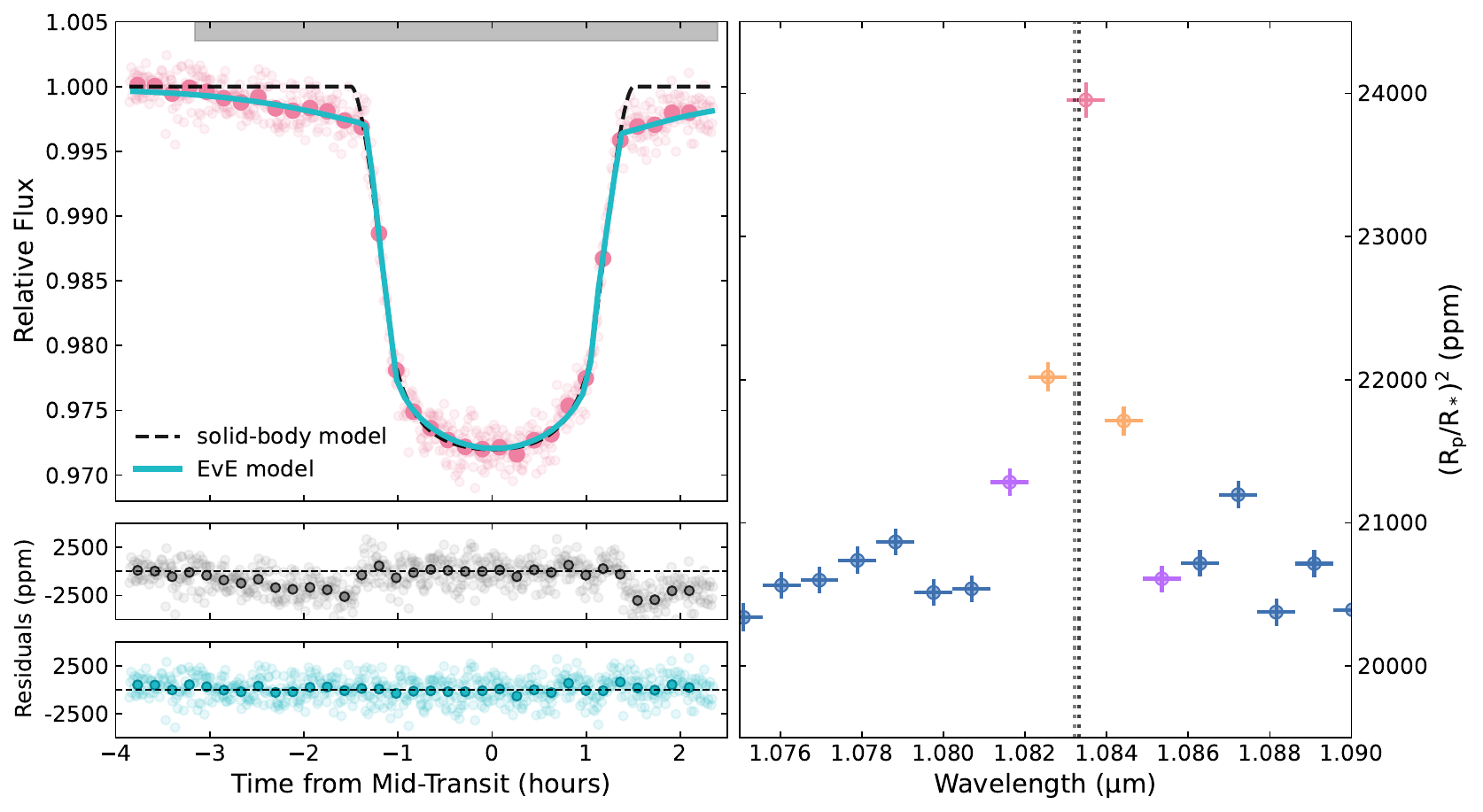}
  \caption{\textit{Left:} Light curve near the metastable helium feature (pixel corresponding to $\lambda$: 1.0830216 -- 1.0839538\,$\mu$m), overlaid with the best-fit solid-body model light curve (black) and the best-fit \texttt{EvE} model (turquoise), along with their corresponding residuals below. The binned data are highlighted for clarity. Pre-transit and post-transit absorption is evident. The out-of-transit excess absorption is particularly apparent in the middle panel, which shows the residuals from the solid-body model fit. The shaded gray region indicates the coverage of past ground-based observations of the target. \textit{Right:} Transmission spectrum at instrument resolution around the metastable helium lines. Although the lines are not fully resolved (indicated by black dotted lines), the excess absorption is clearly visible. The extended thermosphere light curves for the wavelength bins near the metastable helium lines (highlighted in purple, orange, and pink) are shown in Figure~\ref{fig: helium_append}.}
  \label{fig: helium_trans_lc}
\end{figure*}

At the metastable helium wavelength bin (pixel resolution; centered at 1.08348\,$\mu$m), we measure a maximum transit depth of 2.395\,\% $\pm$ 0.01\,\% (36\,$\sigma$). This corresponds to a time-averaged excess transit depth of 0.340\,\% $\pm$ 0.01\,\%, which is consistent with the CARMENES and KECK-NIRSPEC high-resolution excess absorption of $\sim\,5.5\,\%$ \citep{allart2019_helium,kirk2020_helium} (corresponding to $0.40\pm0.07\,\%$ at NIRISS-SOSS resolution) and with the \textit{HST} low-resolution absorption of $0.049\pm0.011\,\%$ \citep{spake2018}. Although the helium triplet lines are not resolved in the SOSS spectra, continuous increased transit depths are clearly observed across four wavelength bins centered around 1.083\,$\mu$m (see Figure~\ref{fig: helium_trans_lc}).

\begin{figure}[]
  \centering
  \includegraphics[width=\linewidth]{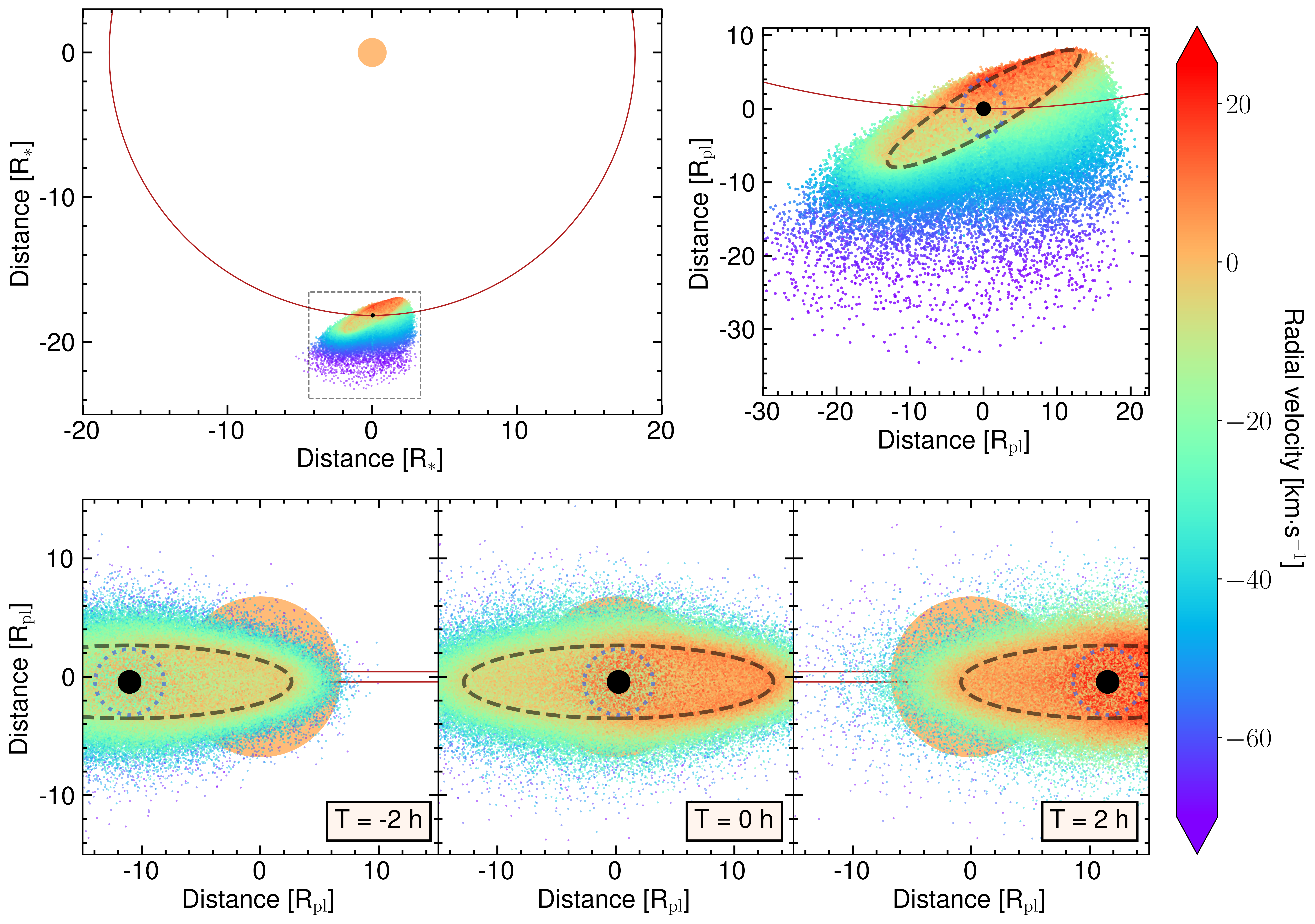}
  \caption{View of the escaping metastable helium in our best-fit model. The red line denotes the planetary orbit, the blue dotted line shows the projection of the Roche lobe, and the black dashed line indicates the boundary of the confined outflow, from which escaping atoms are launched over the 3D surface. The thermospheric profile is generated with a mass-loss of $\sim 10^{12}$\,g/s, a temperature of $7000$\,K and a H/He ratio of 0.90. The associated metastable helium mass-loss is $\sim 9\cdot 10^{5}$\,g/s. \textit{Top}: Views of the system from above at mid-transit. The right panel is a zoom in over the square black region indicated in the left panel. \textit{Bottom}: Views of the system along the line-of-sight at mid-transit, and 2 hours before/after.}
  \label{fig: helium_cloud}
\end{figure}

We report, for the first time, the detection of pre-transit helium absorption from \textit{JWST} at 17\,$\sigma$. While post-transit helium absorption has been observed previously at high resolution \citep{spake2021_helium}, this unprecedented detection of pre-transit absorption along with post-transit (19\,$\sigma$) absorption enables us to reveal a large extension of the planet’s thermosphere (see Figure~\ref{fig: helium_cloud}). The large spatial extent of the helium thermosphere around WASP-107\,b, combined with the high signal-to-noise (S/N) ratio from this continuous SOSS observation, made it possible to detect the pre-transit absorption. It is worth noting that for HAT-P-32\,b \citep{hat-p-32b_helium_science} and HAT-P-67\,b \citep{hatp67_helium}, helium has been previously detected across the full orbital phase using ground-based high-resolution spectroscopy, with observations stitched together from multiple nights. In these very extended streams, the observed escaping helium no longer tracks the planet’s orbital motion but its own dynamics, showing only a slight velocity shift from the stellar rest frame. In contrast, in our continuous SOSS dataset of WASP-107\,b, we observe a pre-transit slope in the wavelength bin centered at 1.08348\,$\mu$m, which we attribute to a super extended helium atmosphere that remains confined close to the planet. The pre-transit increase in absorption begins approximately 1.5 hours before ingress and continues until the end of our observation window, 1 hour after egress. We argue in Section \ref{subsec: helium_methods} that the observed signal cannot be explained by systematics or stellar variability. The pre-transit increase in absorption begins approximately 1.5 hours before ingress and continues until the end of our observation window, 1 hour after egress.

Using the Evaporating Exoplanets (\texttt{EvE}) code \citep{bourrier_2013}, we modeled WASP-107\,b's escaping atmosphere with an ellipsoidal thermosphere, informed by extended exosphere observations \citep{Bourrier2018} and hydrodynamical simulations of atmospheric outflows interacting with stellar winds \citep[e.g.,][]{Wang2021,MacLeod2022}. In fact, this extended absorption duration cannot be explained by a spherical thermosphere confined within the planetary Roche Lobe, even when including an exosphere, given the short lifetime of metastable helium ($\sim$\,2 hours). We simplify the more complex structure predicted by hydrodynamical simulations to estimate the spatial extent of the thermosphere trailing the planet. As shown in Figure~\ref{fig: helium_trans_lc}, our model reproduced the pre-transit slope, constraining the confined outflow's elongation ahead of the planet to 10 -- 18 planetary radii (8.5--15 in line-of-sight projection). The post-transit elongation is similar, with no clear evidence of a shifted ellipsoid center. 


At the NIRISS-SOSS resolution (R\,$\sim$\,700), a degeneracy exists between the line shape and depth of the absorption signal, resolvable only with high-resolution observations. Previous studies of metastable helium on this planet used unocculted stellar spectra constructed during pre- or post-transit phases (grey area in Figure~\ref{fig: helium_trans_lc}), likely biasing their light curves and spectra as we discuss in Section \ref{subsec: atmos_escape}. To avoid this, we excluded these previous observations from our analysis of WASP-107\,b's mass-loss. This underscores the need to coordinate ground- and space-based observations, thereby obtaining a high S/N phase curve from \textit{JWST} combined with the resolved line profile from high-resolution observations to fully model and interpret the helium signature. Notably, our continuous observation is the first to detect both pre- and post-transit absorption, emphasizing the importance of revisiting helium detections with \textit{JWST} observations with longer baselines for more accurate interpretations.

\begin{figure*}[]
    \centering
    \includegraphics[width=\linewidth]{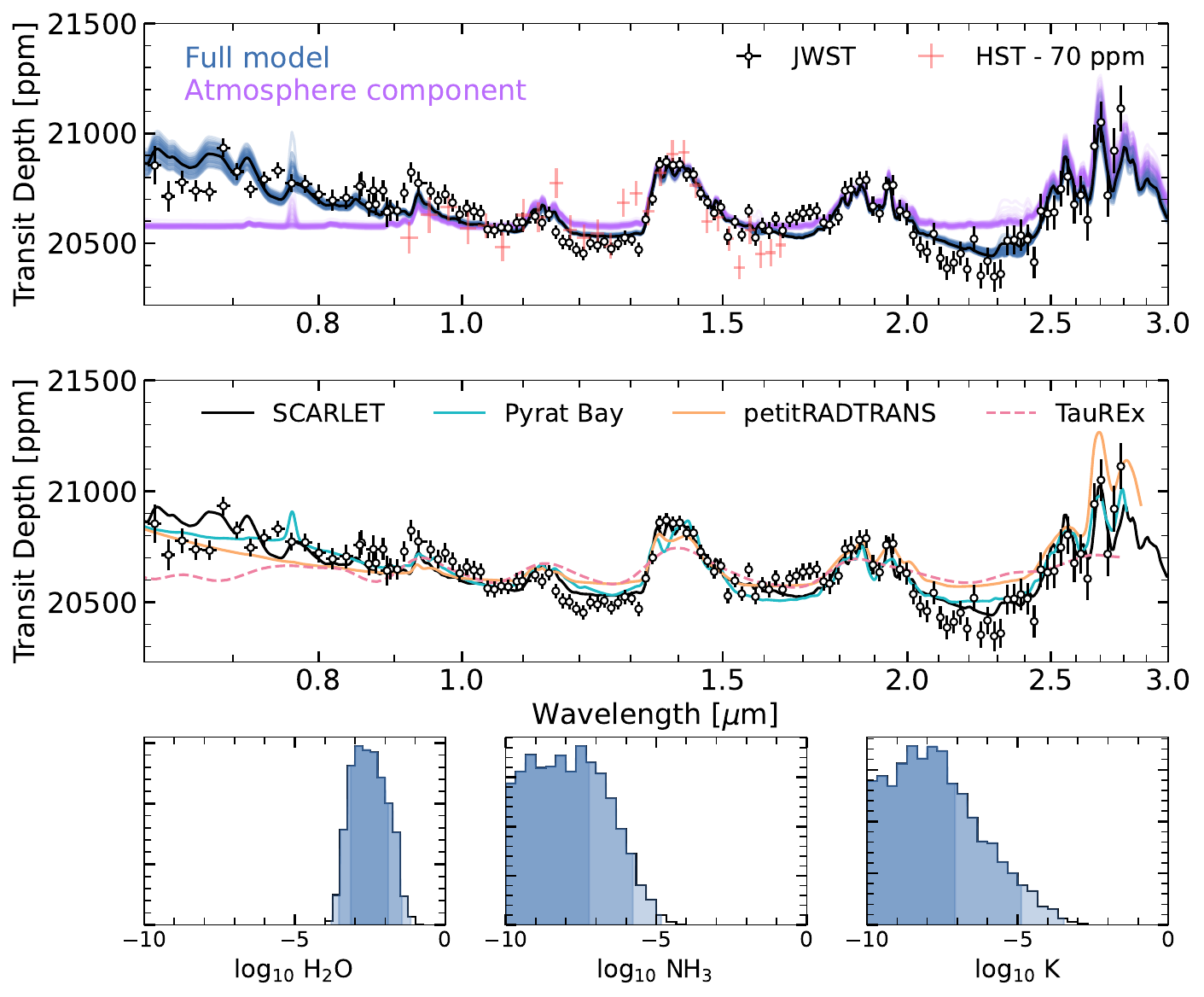}
    \caption{Results of free retrieval performed on the \textit{JWST}/NIRISS-SOSS transmission spectrum of WASP-107\,b with \texttt{SCARLET}, \texttt{TauREx}, \texttt{petitRADTRANS} and \texttt{Pyrat Bay}. \textit{Top panel:} Sample spectra from the posterior distributions of the \texttt{SCARLET} free retrievals (joint fit of planetary atmosphere and stellar contamination). The full models are shown in blue, and the atmosphere contribution is shown in purple for each sample. The best-fit \texttt{SCARLET} model is shown in black. The fitted NIRISS-SOSS transmission spectrum is overlaid (black points) and shifted by the best-fitting offset from the \texttt{SCARLET} retrieval (156 ppm), while the \textit{HST} G102/G141 spectrum is shown in red (shifted by 70 ppm; \citealp{wellbanks2024_nircam}). \textit{Middle panel:} Best-fit spectra retrieved with \texttt{SCARLET} (black), \texttt{TauREx} (dashed pink), \texttt{Pyrat Bay} (blue) and \texttt{petitRADTRANS} (orange). \texttt{TauREx} retrievals poorly fit the data and are not used for interpretation. \textit{Bottom panels:} Posterior distributions on the H$_2$O, NH$_3$ and K volume mixing ratios from the \texttt{SCARLET} retrievals. The shaded areas indicate the 1, 2, and 3$\sigma$ confidence intervals. For K and NH$_3$, where only upper limits are obtained, we show the 1, 2, and 3$\sigma$ upper limits. Upper limits are reported at 2$\sigma$ above the histograms.}
    \label{fig:retrieval_spectrum_fit}
\end{figure*}

Building on these observations, we extended our analysis to the broader transmission spectrum of WASP-107\,b. By combining the detailed helium signal with the continuum and molecular absorption features, we gain a more comprehensive understanding of the planet's atmosphere and stellar environment. We used four different frameworks: \texttt{SCARLET} \citep{benneke_distinguishing_2012,benneke_characterizing_2013,benneke_strict_2015,benneke_sub-neptune_2019,benneke_water_2019,pelletier_where_2021,piaulet_evidence_2023,Piaulet_Ghorayeb_2024}, \texttt{petitRADTRANS} \citep{molliere_prt1}, \texttt{Pyrat Bay} \citep{PyratBay} and \texttt{TauREx} \citep{Taurex} for our retrievals, of which \texttt{SCARLET} could account for stellar surface heterogeneities. Our measurement of a precise and pronounced slope of increased transit depth with decreasing wavelength is best explained by the presence of unocculted stellar spots on the stellar surface (the transit light source effect - TLSE; \citealp{rackham_transit_2018}) rather than only a haze slope ($\Delta \ln Z=11.62$ in favor of the model with heterogeneity, or 5.2$\sigma$). While our spectrum is consistent over the overlapping 0.9--1.7$\mu$m wavelength range with the \textit{HST}/WFC3 G102/G141 spectrum \citep{kreidberg_water_2018,spake_helium_2018,welbanks_high_2024}, the precision we achieve enables us to determine that stellar contamination is favored compared to small-particle haze opacity (see Figure~\ref{fig:retrieval_spectrum_fit}). The wavelength coverage of NIRISS-SOSS, compared to the other \textit{JWST} instruments, is the optimal range when it comes to obtaining a robust estimate of the atmospheric water abundance since it covers multiple prominent water features. 

We find that stellar contamination alters the shape and depth of water features in the atmosphere of WASP-107\,b and caution that joint retrievals accounting for the impact of stellar contamination are necessary to obtain unbiased water abundance estimates for this planet (Figure~\ref{fig:sensitivity_abundances}). The \texttt{SCARLET} joint retrieval measures $\log_{10}$ H$_2$O = $-2.5 \pm 0.6$ (Figure~\ref{fig:posterior_molecules_compare}), robust to our model assumptions (Figure~\ref{fig:sensitivity_abundances}) and mostly shaped by the inclusion of stellar contamination in the model. When running a retrieval without accounting for stellar contamination, we find that the water abundance is overestimated by about a factor of 40 (Figure~\ref{fig:sensitivity_abundances}) as the muted water bands and the blueward slope are fitted with a haze slope and a higher-metallicity atmosphere. 

Additionally, we obtain upper limits on the abundances of other volatile and refractory species where constraints were previously not available or model-dependent because of the lack of short-wavelength spectral coverage (Figure~\ref{fig:posterior_molecules_compare}). While the NIRISS-SOSS spectrum is sensitive to NH$_3$ and K, we do not detect either of these species and obtain upper limits of 1.7 ppm (for NH$_3$) and 13 ppm (for K) on their atmospheric abundances, at the 2$\sigma$ level. The low NH$_3$ abundance we measure is consistent with the upper-atmosphere depletion expected from vigourous vertical mixing, which is expected to be fueled by the high interior temperature of WASP-107\,b \citep{sing_warm_2024,welbanks_high_2024}. The upper limit we obtain on the K abundance enables us to infer a K/H$<75 \times$ stellar at 2$\sigma$, consistent with the enhanced metallicity of $8^{+26}_{-6}\times$\,stellar we derive from the \texttt{SCARLET} measurement of the H$_2$O abundance. 
The planet's atmospheric metallicity is slightly enhanced relative to (but compatible with) the O/H mass-metallicity trend observed for exoplanets (\citet{welbanks_degeneracies_2019}; Figure~\ref{fig:mass_metallicity}) which reflects the impact of core accretion on atmospheric compositions \citep{thorngren_mass-metallicity_2016}.

Nearly two-thirds of WASP-107\,b's mass is predicted to reside in its atmosphere \citep{sing2024_nirspec}. From our inferred super-solar water metallicity ($8^{+26}_{-6}\times$\,stellar) and the order of magnitude mass-loss rate derived from the helium thermospheric profile ($\sim$\,1--10\,M$_\oplus$/Gyr), we estimate that the planet had a metallicity of 4\,$\times$\,stellar at birth, if it formed in-situ at the current orbit. For a close-in giant planet orbiting a near-solar metallicity star ([Fe/H] = +0.02; \citep{wasp107b_caroline2021}), the accretion of high metalicity gases is highly unlikely \citep{2019A&A...627A.127C}, making the in situ formation improbable. 

One plausible scenario for WASP-107\,b's evolution is that it formed further out accreting gases and refractories from the protoplanetary disk \citep{formation_ctoo}. It might have migrated inward, potentially due to interactions with a third body in the system \citep[WASP-107c;][]{wasp107b_caroline2021}. Its non-circularized orbit (e = 0.06\,$\pm$\,0.04) and polar retrograde orientation \citep{wasp107b_daiwinn2017, rubenzahl2021_rossiter} strongly support this migration scenario.

As the planet is currently undergoing circularization, strong tidal forces in WASP-107b are heating its interior \citep{2002AREPS..30..113B, 2008ApJ...681.1631J, 2019ApJ...886...72M}, as supported by the high internal temperatures observed with NIRSpec+MIRI \citep{sing2024_nirspec} and NIRCam+MIRI \citep{wellbanks2024_nircam}. Tidal heating can explain the planet’s inflated atmosphere, which may enhance upper-atmosphere escape and sustain the outflow detected in our helium observations. Alternatively, Ohmic dissipation could also account for the high internal heat, removing the need for tidal heating as the primary source \citep{tidal_wasp107b_2025}. While this reduces the requirement for recent migration (i.e., within the past $\lesssim$\,2\,Myr), it does not rule out the possibility that WASP-107\,b has undergone migration.


Our continuous SOSS observations reveal the presence of an extended thermosphere on either side of the planet and confirm that the planet is sustaining an outflow. This time-series transit observation is the first of its kind to detect pre-transit helium absorption from space, which remained undetected from the ground \citep{allart_high-resolution_2019, kirk2020_helium, spake_posttransit_2021} for WASP-107\,b. This suggests that short-baseline helium observations may underestimate the full extent of exoplanetary thermospheres. Revisiting other helium-detected planets with extended time-series observations is therefore essential, either to assess time variability or to improve the precision and completeness of their light curves.

\section{Methods} \label{sec: methods}

Transmission spectroscopy has been the primary tool for exploring exoplanet atmospheres \citep{seagersasselov2000, charbonneau2002}. When a planet transits in front of its star some of the starlight passing through the planet's atmosphere is absorbed. These absorption signatures reveal the atomic and molecular composition of the atmosphere.  In turn, this information sheds light on the formation and evolutionary history of the planet \citep{formation_ctoo}. 

A key process in the evolution of a planet is atmospheric escape \citep{Tian2005, Lammer2008, yalle2008}, driven by stellar high-energy X-ray and extreme-ultraviolet (XUV) radiation \citep{Lammer2003, owenjackson2012}, or by the cooling core of a hot planet \citep{ginzburg2018}. Absorption and emission lines originating from electronic transitions in hydrogen and helium are the primary means of capturing atmospheric escape, along with metal lines \citep{seagersasselov2000}. One such marker is the metastable helium line at 1.083\,$\mu$m \citep{oklopcichirata2018, oklopcic2019}. This line is unique because it is not affected by the interstellar medium (ISM) \citep[unlike the Lyman-$\alpha$ line, e.g.,][]{lymana_vidal2003}, and can be observed from both space-based \citep[e.g.,][]{spake2018} and ground-based telescopes \citep[e.g.,][]{allart2019_helium}.

In addition to hydrogen and helium, the abundance of trace gases offer valuable insights into the underlying physical and chemical processes in the atmosphere. These processes can be interpreted through carbon and oxygen chemistry, revealing various formation pathways and evolutionary histories \citep[e.g.,][]{boucher2023_wasp127b_spirou}. However, it is challenging to measure the abundances of all the carbon- and oxygen-bearing species to constrain the C/O ratio and metallicity \citep{Tsai2021}. Short-period, low-density giant planets are ideal targets for this. Their high temperatures keep most of the C- and O-bearing molecules in vapor phase and contribute to large atmospheric scale heights. This, in turn, opens the door for constraining the abundances through transmission spectroscopy. 

\subsection{About WASP-107\,b} \label{subsec: about_w107b}

WASP-107\,b \citep{wasp107b_anderson2017} is an ultra-low density super-Neptune (0.96 $\pm$ 0.03\,$\mathrm{R_{J}}$) orbiting a K6 star. It is an ideal target for constraining the primordial H/He-dominated atmosphere. \citet{wasp107b_caroline2021} estimates the mass to be 30.5 $\pm$ 1.7\,$\mathrm{M_{\oplus}}$, placing the planet right in the runaway gas accretion regime without becoming a gas giant. They also argued that the planet's core-mass of $<$\,4.6\,$\mathrm{M_{\oplus}}$, substantial volatile envelope, and non-zero eccentricity ($\sim$\,0.06), indicates that the planet might have migrated inward to the inner disk. The Rossiter–McLaughlin (RM) measurements of WASP-107\,b \citep{wasp107b_daiwinn2017, rubenzahl2021_rossiter}, showed a retrograde, polar orbit for the planet, again suggesting the planet migrated to its current orbit. 

Using \textit{HST}, \citet{kreidberg2018_water_hst} identified a super-solar (30$\times$\,Solar) enriched atmosphere on WASP-107\,b, characterized by water and depleted methane. The planet was observed in transit with \textit{JWST}'s NIRSpec \citep{sing2024_nirspec}, MIRI \citep{achrene2024_miri}, and NIRCam \citep{wellbanks2024_nircam}. All three spectra are consistent with a metal-enriched atmosphere containing sulfur dioxide, high internal temperatures but depleted in methane. However, \citet{sing2024_nirspec} predict a higher core mass fraction (11.5\,$\mathrm{M_{\oplus}}$), supporting the core-accretion model.


\subsection{Data reduction and light curve fitting} \label{subsec: data_reduction}

We reduced the \textit{JWST}/NIRISS-SOSS TSO of WASP-107\,b with the open source pipeline \texttt{exoTEDRF} \citep{exotedrf, radica2023_wasp96b, feinstein2023_wasp39b}, starting with the raw uncalibrated data products from the MAST archive. For data reduction, we apply stages 1, 2, and 3 of \texttt{exoTEDRF} closely following the procedure laid out in \citep{Radica2024, Cadieux2024, Benneke2024}. More specifically, we correct the column-correlated 1/$f$ noise at the group level (before ramp fitting), and use a piecewise background subtraction (independently scaling the STScI SOSS background model either side of the background ``step'' \citep{Lim2023, Fournier-Tondreau2024}). We extract the time-dependent stellar spectra using a simple box aperture with a width of 32 pixels, as the order-self-contamination is negligible for this target over the wavelength range that we consider here (i.e., 0.6--0.85\,µm for order 2 and 0.85 -- 2.85\,µm for order 1) \citep{Darveau-Bernier2022, Radica2022}. In fact, WASP-107\,b actually has the largest predicted level of order self-contamination of the 13 targets considered by \citep{Darveau-Bernier2022}, reaching up to $\sim$500\,ppm at $\sim$1.1\,µm in order 2. However, as is generally done, we disregard wavelengths $>$0.85\,µm in order 2, and the order self-contamination level is restricted to $<$10\,ppm over the wavelength range we consider -- smaller than our fitted error bars. There are four order 0 contaminants that intersect the target spectral trace and contaminate the extracted spectra \citep[e.g.,][]{radica2023_wasp96b}. These contaminants are located at the following wavelengths: 0.935--0.945\,µm, 1.098--1.112\,µm, 1.775--1.785\,µm, and 1.834--1.844\,µm, and are masked in the remainder of our analysis. 

\begin{figure*}[]
  \centering
  \includegraphics[width=\linewidth]{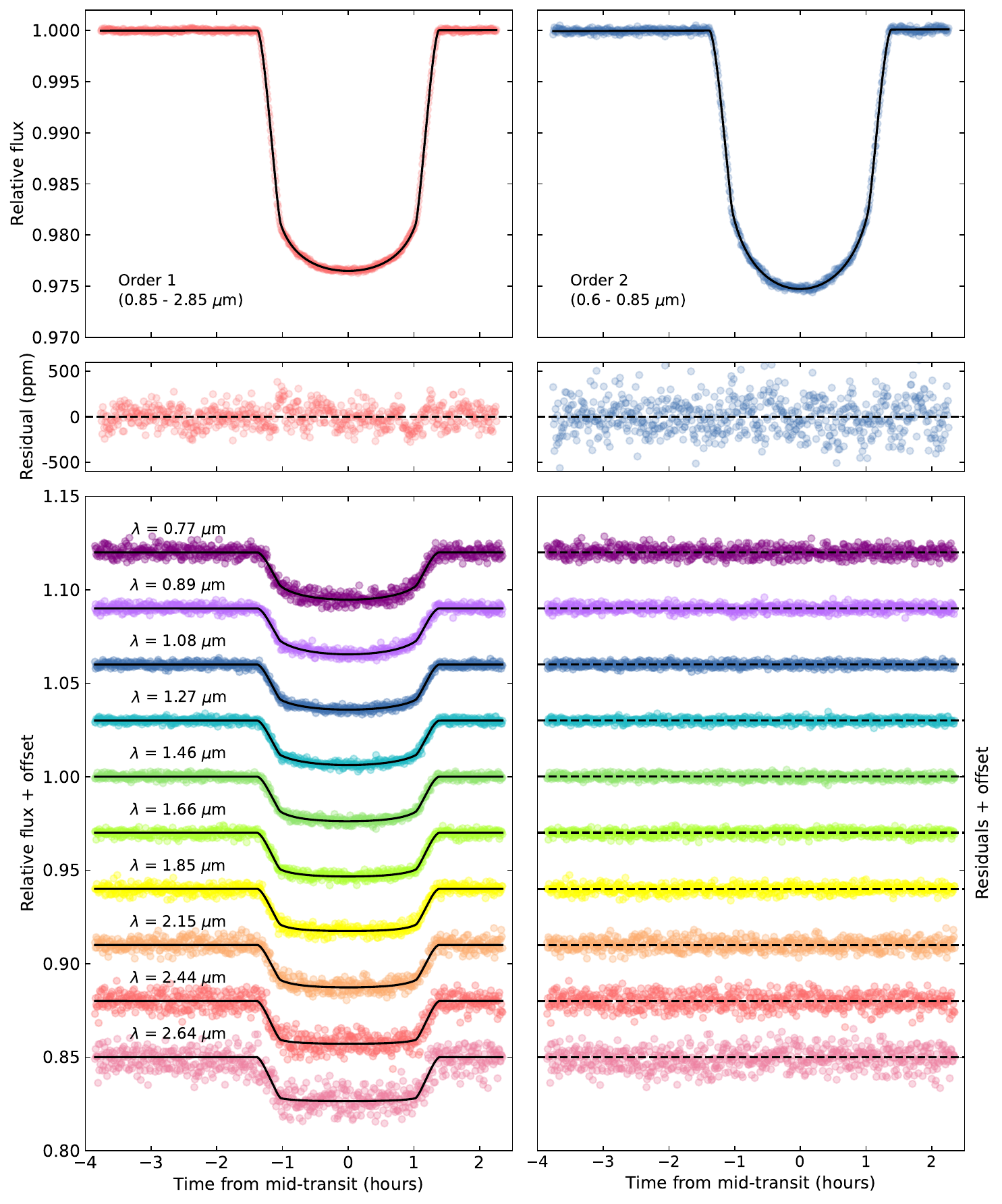}
  \caption{\emph{Top:} Broadband light curves of NIRISS-SOSS order 1 (red) \& 2 (blue) and best-fit transit models (black). \emph{Middle:} The corresponding residuals. \emph{Bottom:} Representative instrumental resolution light curves with their residuals on the right. No systematic trends have been removed from the data.}
  \label{fig: light_curves}
\end{figure*}

We binned all wavelengths from orders 1 and 2 separately to create broadband light curves. These broadband light curves were then fitted using \texttt{juliet} \citep{juliet}, which utilizes transit light curves generated by \texttt{batman} \citep{batman}. We fitted for mid-transit time, impact parameter and limb-darkening coefficients, while keeping eccentricity, orbital period and argument of periastron fixed. We sampled the posteriors with 1000 live points using the \texttt{dynesty} nested sampling package \citep{dynesty} (see top panels in Figure~\ref{fig: light_curves}). 

The mid-transit time (T$_0$) and impact parameter (b) from each of the broadband light curves were extracted and fixed when fitting the pixel-resolution time series light curves using \texttt{juliet}. The quadratic limb-darkening parameters and transit depths were derived from these fits. No detrending was performed while fitting the pixel-resolution spectroscopic time series, as the broadband fits did not exhibit any significant trends (see Section \ref{lc_fits_append} for more information on the de-trending fits). Representative pixel-resolution light curves and their fit residuals are shown in the bottom panels of Figure~\ref{fig: light_curves}. The stellar and planetary parameters used in our analysis, as well as those retrieved from it, are summarized in Table~\ref{table:params}. 

For robustness, we reduce the data using another independent pipeline: \texttt{NAMELESS} \citep{coulombe2023_nameless_wasp18b, feinstein2023_wasp39b}. The transmission spectra we get from both pipelines are virtually identical (0.76\,$\mathrm{\sigma}$ mean deviation; Figure~\ref{fig: comparison}). We used \texttt{exoTEDRF} data products for our analyses. The reduction procedure using the \texttt{NAMELESS} pipeline and \texttt{ExoTEP} light curve fitting are detailed in Appendix \ref{data_red_appendix}. Additionally, we performed retrievals on the \texttt{NAMELESS} dataset, which yielded results comparable to those from our \texttt{exoTEDRF} dataset. These results are presented in Appendix \ref{data_red_appendix} and in Figure \ref{fig: nameless_retrieval}.

\begin{table*}[t]
\begin{minipage}[tbh!]{\textwidth}
\caption{Stellar and planetary parameters of the WASP-107 system}
\label{table:params}
\small
\centering
\begin{tabular}{l l c l}
\hline
\hline
Parameter & Symbol [Unit] & Value & Source \\
\hline
\hline

\multicolumn{4}{l}{Stellar Parameters} \\
Stellar mass & $M_\mathrm{\star}$ [$M_\mathrm{\odot}$] & $0.683_{-0.016}^{+0.017}$ & \cite{wasp107b_caroline2021}\\
Stellar radius & $R_\mathrm{\star}$ [$R_\mathrm{\odot}$] & 0.67 $\pm$ 0.02 & \cite{wasp107b_caroline2021}\\ 
Effective temperature & $T_\mathrm{eff}$ [K] & 4425 $\pm$ 70 & \cite{wasp107b_caroline2021}\\
Metallicity & [Fe/H] [dex] & +0.02 $\pm$ 0.09 & \cite{wasp107b_caroline2021} \\
Surface gravity & $log\,g_\mathrm{\star}$ [cgs] &  4.633 $\pm$ 0.012 & \cite{wasp107b_caroline2021}\\
Spectral type & & K6 & \cite{wasp107b_anderson2017} \\
Limb-darkening (SOSS O1) & $u_\mathrm{1}$ & 0.2723$^{+0.0205}_{-0.0177}$ & This work \\
Limb-darkening (SOSS O1) & $u_\mathrm{2}$ & 0.2098$^{+0.0106}_{-0.0055}$ & This work \\
Limb-darkening (SOSS O2) & $u_\mathrm{1}$ & 0.4746$^{+0.0418}_{-0.0346}$ & This work \\
Limb-darkening (SOSS O2) & $u_\mathrm{2}$ & 0.2013$^{+0.0227}_{-0.0138}$ & This work \\
Rotation period & $P_\mathrm{rot}$ [days] & 17 $\pm$ 1  & \cite{wasp107b_anderson2017} \\
Stellar age & Age [Gyr] & 3.4 $\pm$ 0.7 & \cite{wasp107b_caroline2021}\\
\hline

\multicolumn{4}{l}{Planetary Parameters - WASP-107\,b} \\
Planet mass & $M_\mathrm{p}$ [$M_{\oplus}$] & 30.5 $\pm$ 1.7 & \cite{wasp107b_caroline2021} \\
Planet radius & $R_\mathrm{p}$ [$R_{J}$] & 0.96 $\pm$ 0.03 & \cite{wasp107b_caroline2021}\\
Equilibrium temperature & $T_\mathrm{eq}$ [K] & 770 $\pm$ 60 & \cite{wasp107b_anderson2017} \\
Mid-transit epoch (SOSS O1) & $T_\mathrm{c}$ [$BJD_\mathrm{TDB}$]& 2460107.50587072$^{+0.00000871}_{-0.00000992}$ & This work \\ 
Mid-transit epoch (SOSS O2) & $T_\mathrm{c}$ [$BJD_\mathrm{TDB}$]& 2460107.50585112$^{+0.00001922}_{-0.00001596}$ & This work \\
Orbital period & P [d] & 5.7214742 $\pm$ 0.0000043 & \cite{wasp107b_daiwinn2017}   \\
Orbital semi-major axis & a [AU] & 0.0566 $\pm$ 0.0017 & \cite{wasp107b_caroline2021} \\
Orbital inclination & i [deg] & 89.8 $\pm$ 0.20 & \cite{wasp107b_daiwinn2017} \\
Impact parameter (SOSS O1) & b & 0.1191$^{+0.0167}_{-0.0216}$ & This work \\
Impact parameter (SOSS O2) & b & 0.1395$^{+0.0236}_{-0.0364}$ & This work \\
Eccentricity & e & 0.06 $\pm$ 0.04 & \cite{wasp107b_caroline2021}\\
Argument of periastron & $\omega$ [deg] & 40$^{+40}_{-60}$ & \cite{wasp107b_caroline2021}\\

\hline
\end{tabular}
\end{minipage}
\end{table*}

\subsection{Helium analysis} \label{subsec: helium_methods}

Observations with \textit{HST} \citep{spake2018} and ground-based high-resolution spectroscopy \citep{allart2019_helium, kirk2020_helium, spake2021_helium} have demonstrated that WASP-107\,b possesses an extended helium outflow with post-transit absorption. Our observations reveal additional pre-transit absorption, beginning 1.5 hours before ingress, alongside the previously reported post-transit absorption. Unlike the multi-night, discontinuous detection of helium in HAT-P-32\,b \citep{hat-p-32b_helium_science}, continuous ground-based observations of WASP-107\,b did not report any pre-transit absorption. This underscores the necessity of revisiting helium-detected planets with \textit{JWST} to robustly understand the geometry of escaping helium \citep{KrishnamurthyCowan_2024}. For context, early UV ingress has been reported for WASP-12\,b \citep{wasp12_haswell_uv}, GJ 436\,b \citep{gj436b_lya}, and GJ 3470\,b \citep{gj3470b_lya}. While the absorption observed in GJ 436\,b and GJ 3470\,b is attributed to the planets' extended, escaping exospheres, the early ingress in WASP-12\,b has been interpreted as resulting from an accretion stream or a bow shock ahead of the planet, formed by the interaction of its magnetosphere or extended atmosphere with the stellar wind due to its supersonic orbital motion \citep{wasp12_haswell_uv}.

\begin{figure*}[]
  \centering
  \includegraphics[width=\linewidth]{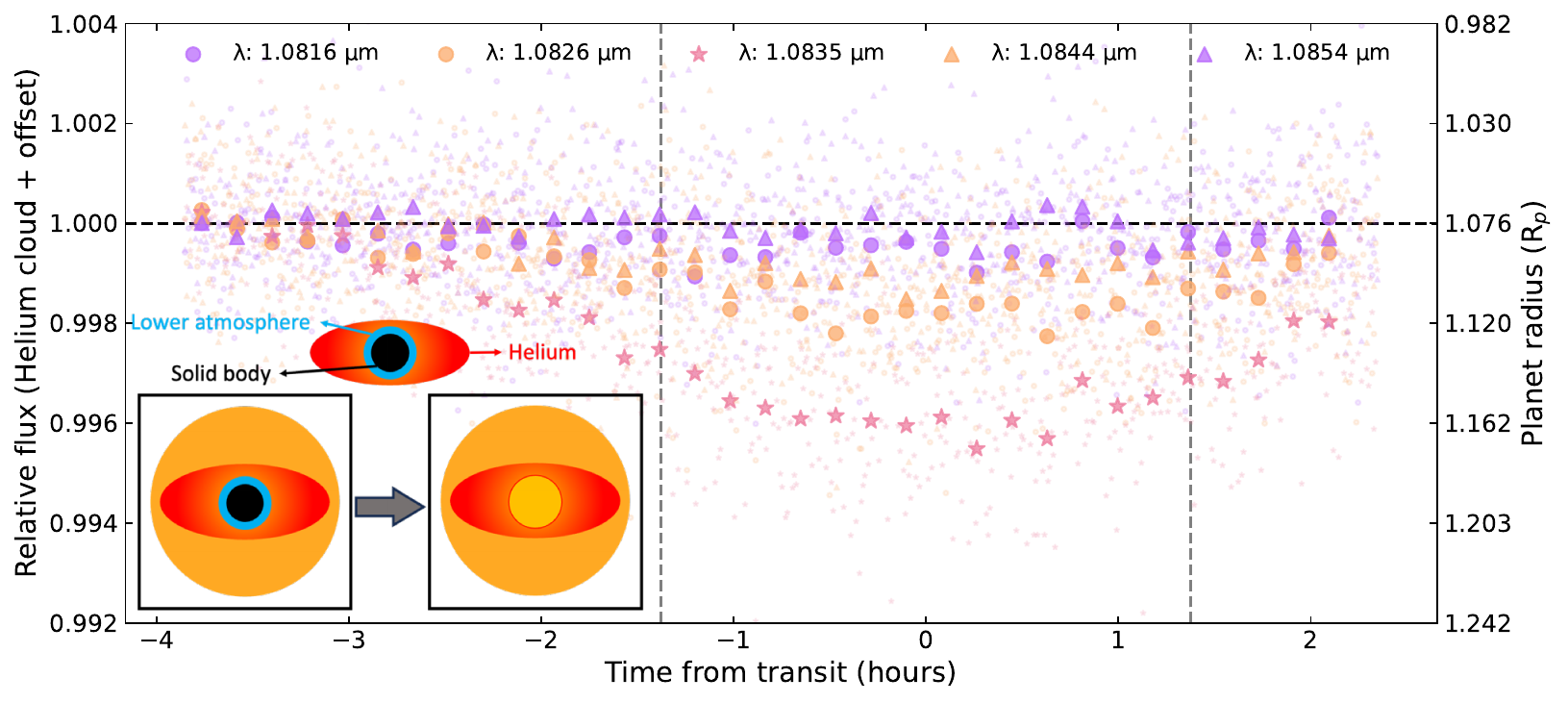}
  \caption{Helium thermospheric light curves (pink stars, orange circles, and orange triangles), obtained after subtracting neighboring wavelength bins, are shown. Our atmospheric model is depicted in the bottom left, with a representative ellipsoidal shape of the helium cloud included for clarity. Pre-transit and post-transit absorption in the wavelength bin centered at 1.0835\,$\mu$m are clearly visible. The helium line (pink stars) maintains baseline flux values until T$_0<$\,3\,hours.}
  \label{fig: helium_append}
\end{figure*}

To understand the origin of the pre-transit slope, we extracted the helium thermosphere light curve by subtracting the average absorption over neighboring wavelength bins around the helium lines, 1070--1080\,nm and 1090--1100\,nm. This effectively removes the contributions from the planet's solid body and lower atmosphere. Figure~\ref{fig: helium_append} shows the light curves of just the helium outflow. By using this, we construct the baseline of the light curves (i.e., $<$\,T$_{\rm 0}$ - 3 hours as our out of transit baseline for helium). Our atmospheric model is depicted in the bottom left of the plot. The excess absorption is still seen in the center wavelength bin of the helium light curve, while we see a flat line in the furthest bins. This indicates that stellar activity alone cannot explain the pre-transit absorption, consistent with an extension of the planet’s outflow between 10 and 18 planetary radii. Using the approach of \citet{spake_helium_2018}, we estimate that unocculted stellar helium absorption could contribute at most one-fourth of the observed feature in our data. This upper limit, based on a 0.4\,Å equivalent width within our 9.3\,Å spectral bin, indicates that the signal cannot be fully explained by stellar activity alone. However, stellar activity can affect the depth of the absorption signal. In order to efficiently remove the stellar contribution from spots and faculae, simultaneous high-resolution observations are needed in activity tracers \citep[see e.g.,][]{guilluy2020hd189helium}. As stellar activity has a small impact on upper atmospheric parameters (Mercier et al. 2025 - accepted, not published, private comms.), we neglect the stellar activity contribution to derive the shape of the planetary outflow.

\subsection{Helium modeling} \label{sebsec: helium_modeling}


We generated synthetic WASP-107\,b transit observations using the Evaporating Exoplanets (\texttt{EvE}) code \citep{bourrier_2013,bourrier_2016}, which simulates planetary transits accounting for geometric and atmospheric effects. We produced a time-series of stellar disk-integrated spectra at high spectral and temporal cadence. These spectra were then convolved with the NIRISS-SOSS instrumental response, averaged over the instrument's spectral grid, and normalized to the out-of-transit spectrum (master-out). To increase the S/N ratio, we temporally resampled the observed NIRISS-SOSS transmission spectra into 5-minute windows and compared with our simulated transmission spectrum. Due to NIRISS-SOSS's low spectral resolution, the depth and shape of absorption features are strongly degenerate. This translates to a degeneracy between mass-loss rate and upper atmospheric temperature in our models.

\begin{figure*}[]
  \centering
  \includegraphics[width=\linewidth]{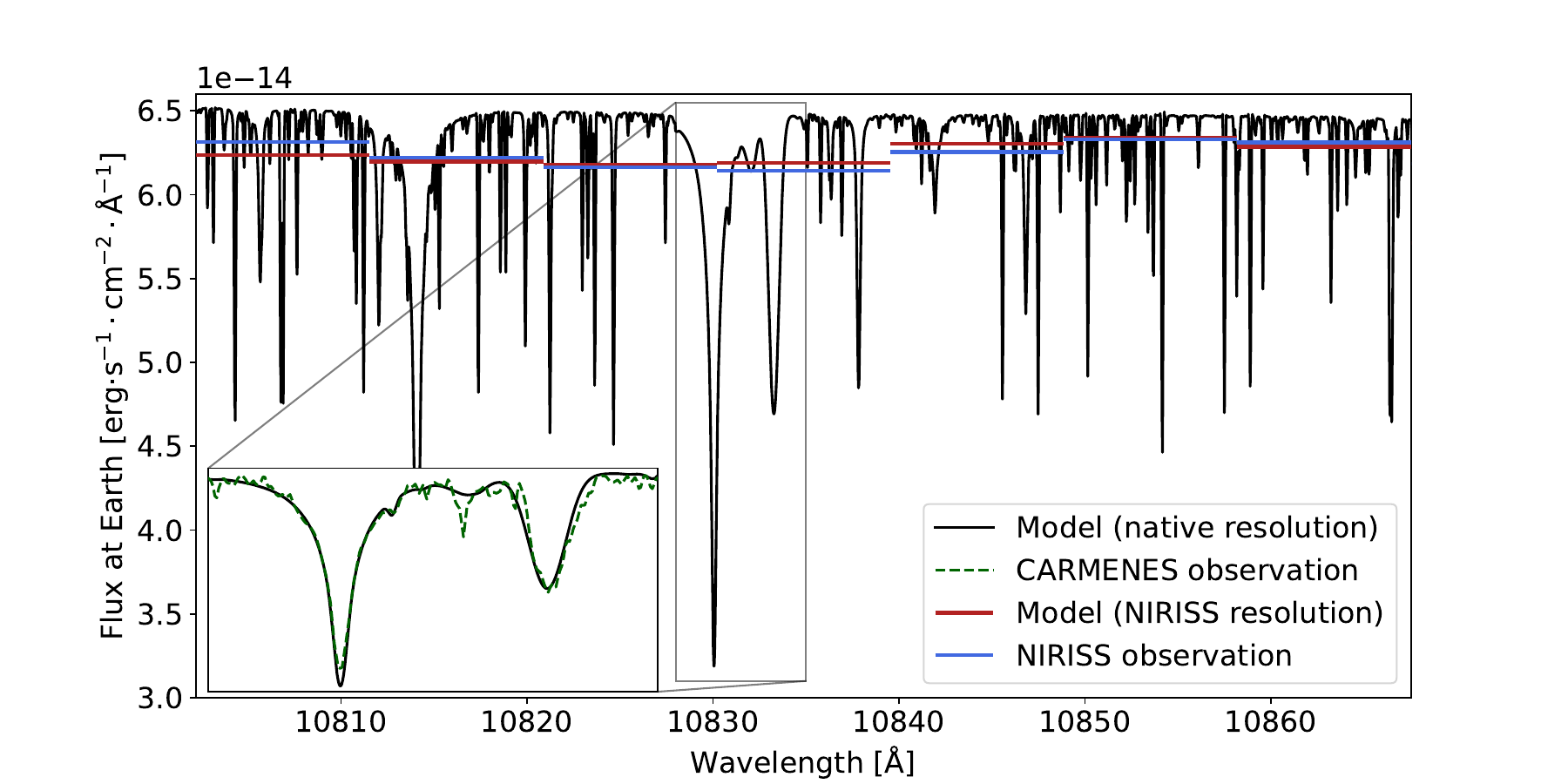}
  \caption{Master-out model used for the helium modeling. We built our model spectrum by combining CARMENES observations and $\textit{Turbospectrum}$ simulations. The zoom-in window corresponds to the wavelength range where the stellar spectrum is synthetic, while outside it corresponds to the CARMENES observations of \citet{allart2019_helium}. We compare our reconstruction to the disk-integrated stellar spectrum as seen by NIRISS-SOSS to adjust the level of the continuum.}
  \label{fig: helium_master_out}
\end{figure*}

\begin{figure*}[]
    \centering
    \includegraphics[width=0.8\linewidth]{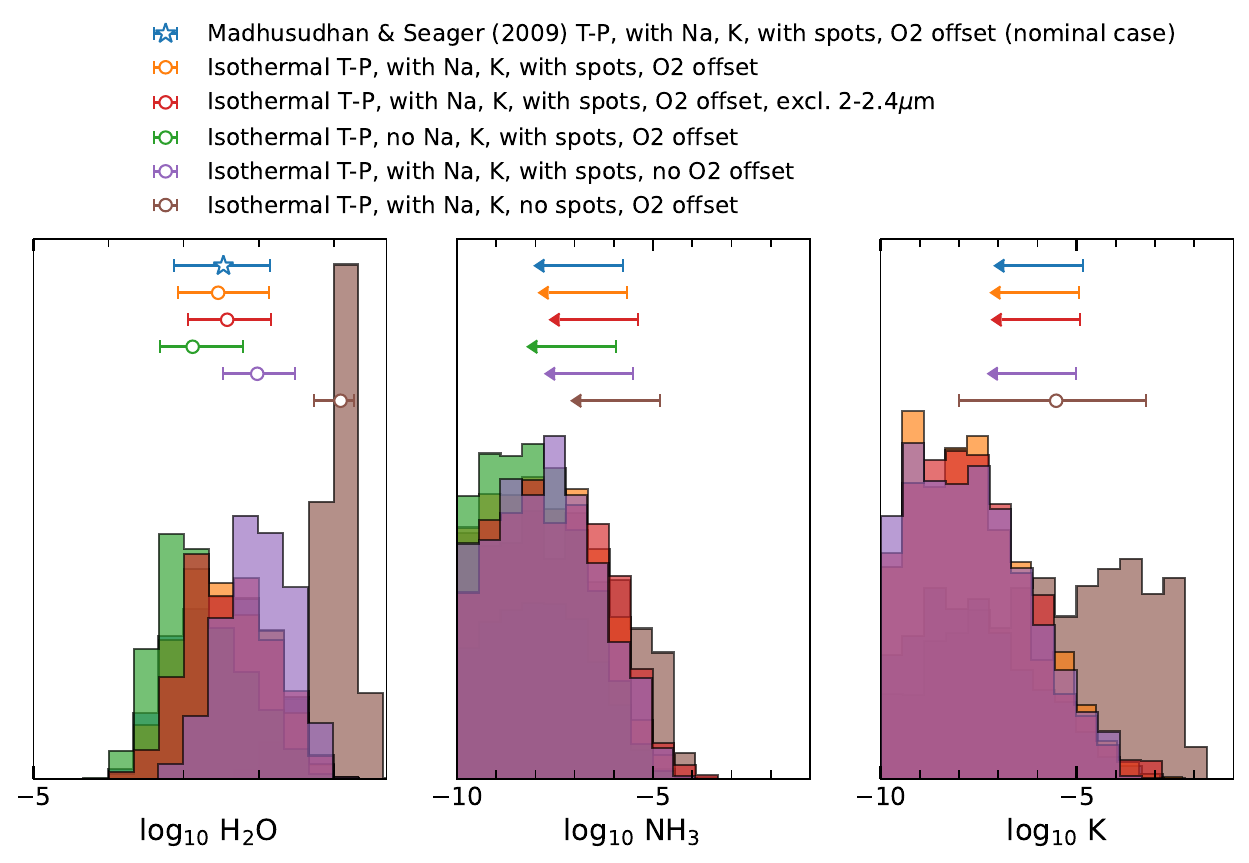}
    \caption{Sensitivity of our NIRISS-SOSS abundance constraints to the retrieval assumptions, explored with the \texttt{SCARLET} retrieval framework. We show the posterior distributions on the abundances of H$_2$O, NH$_3$ and K, as well as how they are impacted by different model assumptions. The 1-$\sigma$ range of water abundances we retrieve, and the 2$\sigma$ upper limits on the NH$_3$ and K abundances, are also shown for each model. We find that our results are not impacted by our choice of T-P profile (isothermal or following the \citet{madhusudhan_temperature_2009} parameterization), the inclusion of the 2-2.4$\mu$m region in the fitted spectrum, and the consideration of Na and K in the set of fitted molecular abundances. Although we retrieve a non-zero order 2 offset with our nominal model (blue, star marker), the inclusion of the offset in the model does not significantly impact the results. We find, however, that neglecting stellar contamination leads to unrealistically high water abundances and an unconstrained potassium abundance.}
    \label{fig:sensitivity_abundances}
\end{figure*}

\begin{figure*}[]
    \centering
    \includegraphics[width=0.8\linewidth]{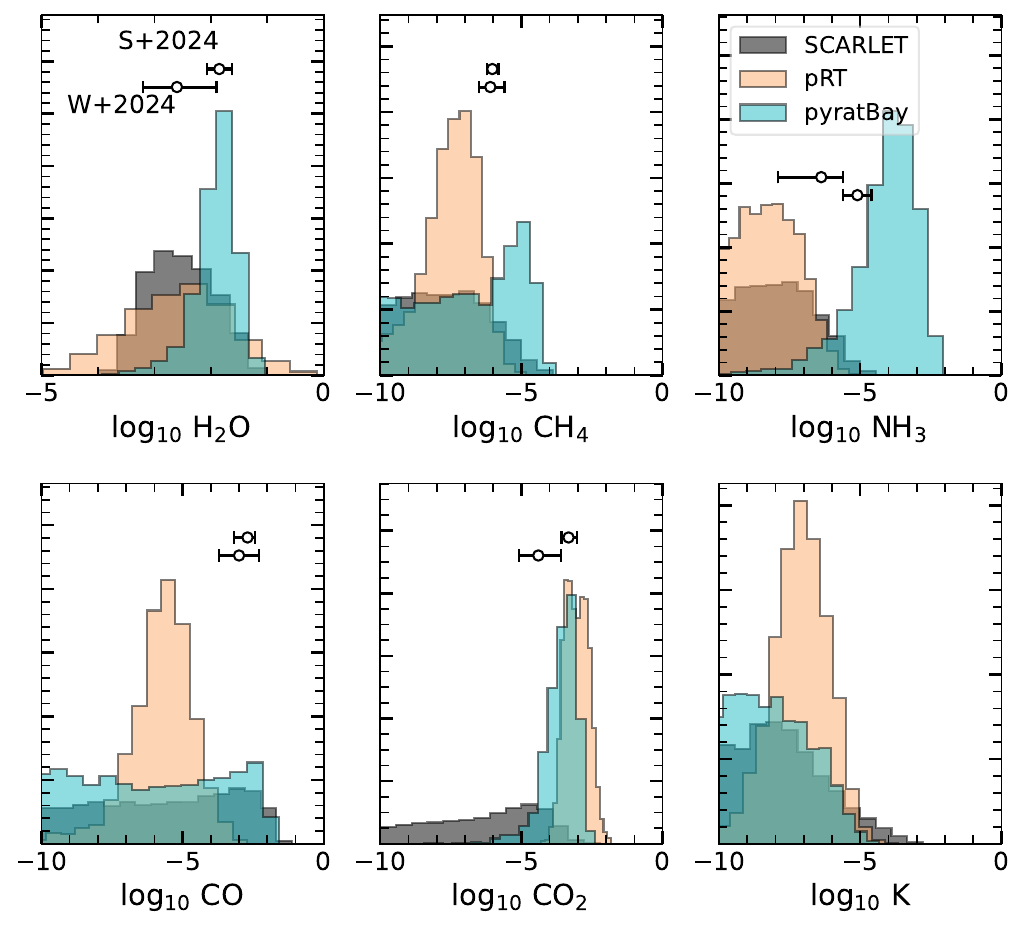}
    \caption{Comparison of the posterior constraints on the molecular abundances from \texttt{SCARLET}, \texttt{Pyrat Bay}, and \texttt{petitRADTRANS} retrievals. The colors match Figure~\ref{fig:retrieval_spectrum_fit} for each retrieval framework. For \texttt{SCARLET}, we show posterior distributions for the fiducial retrieval model. For \texttt{Pyrat Bay} and \texttt{petitRADTRANS}, we show the posterior distributions from the ``definite'' model (sec Section \ref{subsec: atmos_retrieve}) which does not account for stellar contamination or fit an offset between order 1 and order 2. We show the 1$\sigma$ limits on the retrieved abundances of each molecule from \citet{welbanks_high_2024} (labeled as ``W+2024'') in orange (CHIMERA retrieval) and in blue (Aurora retrieval), and from \citet{sing2024_nirspec} (labeled as ``S+2024'', in black). We retrieve consistent abundances with previous literature values, and additionally place an upper limit on the potassium abundance.}
    \label{fig:posterior_molecules_compare}
\end{figure*}

At high spectral resolution, a variety of effects such as the Rossiter-McLaughlin (RM) effect \citep{rossiter_detection_1924,mclaughlin_results_1924,Dethier2023,Carteret2024}, the center-to-limb variations \citep{vernazza_structure_1981,allende_prieto_center--limb_2004,Yan2017} or parameterization of limb darkening \citep{knutson_map_2007} can affect the shape and position of the stellar lines  across the stellar surface. As a result, the local spectra occulted by the planet along the transit chord may not be representative of the disk-integrated stellar spectrum. We model these effects in our analysis to avoid potential biases. We modeled the stellar spectrum using the \textit{Turbospectrum} code\footnote{Latest version, available for download at \href{https://github.com/bertrandplez/Turbospectrum_NLTE/tree/master}{Turbospectrum$\_$NLTE}.} \citep{plez_turbospectrum_2012} and associated line lists \citep{heiter_atomic_2021,magg_observational_2022}. In practice, we generated a stellar grid in which each cell is associated to a local spectrum \citep[based on \textit{Turbospectrum}, see e.g.][]{Dethier2023} that includes all the aforementioned contamination sources. The stellar grid is built using the stellar parameters summarized in Table~\ref{table:params} and fitting the abundance of chemical elements that produce strong stellar lines close to the helium triplet. Because the helium triplet originates from the chromosphere \citep{Vincenzo1997} it is not included in photospheric models such as \textit{Turbospectrum}. We used an analytical model\footnote{Original code from W. Dethier (priv. communication)} that computes the stellar helium triplet using Gaussian profiles for a given helium column density and temperature in the stellar chromosphere. The resulting disk-integrated spectrum is then compared to the CARMENES stellar spectrum in the range 10828--10835\,$\mathrm{\AA}$ \citep{allart2019_helium}. However, since the resolution of NIRISS-SOSS is R\,$\sim$\,700, the convolution kernel is large and the stellar spectrum is needed outside this range to compute the flux time-series. For this, we used the CARMENES disk-integrated spectrum in which we added only the limb darkening effects since the RM effect is negligible at low resolution \citep{Carteret2024}. Finally, we adjusted the level of flux to the NIRISS-SOSS observed level. We show in Figure~\ref{fig: helium_master_out} the disk-integrated spectrum used in the modeling compared to the observed one. We notice small deviations between the reconstructed spectrum and the NIRISS-SOSS data. However, since we fit for the time-series of absorption, we should not be affected by these tiny differences.

The pre-transit slope indicates the presence of an outflow in leading the planet in its orbit. This is consistent with hydrodynamical simulations for a moderate stellar wind strength (see Figure~2 in \cite{MacLeod2022} for instance). We built a model of the expanding upper atmosphere of WASP-107\,b based on previous observations of extended exospheres \citep{Bourrier2018} to account for the stellar wind interaction and to recreate the pre-transit absorption. In our model, we simulated the shape of the atmosphere as an ellipse that is rotated in the orbital plane (Figure~\ref{fig: helium_cloud}). Using a Parker-wind approximation, we generated the density profile of metastable helium with \textit{p-winds} \citep{DosSantos2022}. We set the boundary between the thermosphere and exosphere at the interface of the extended Parker-wind, beyond which particles are in a collisionless regime. We noticed that the nominal level of stellar XUV and bolometric flux does not permit the formation of a smooth tail structure. Indeed, the helium cometary tail was very short-lived due to the high level of photoionization. Previous ground-based observations hinted at the presence of an extended particle regime surrounding the planet, evidenced by deviations from a Gaussian-like absorption spectrum. Similarly to \cite{allart2019_helium}, we reduced the stellar flux by a factor of 50 to facilitate the formation of a smooth tail. However, we note that this choice does not affect our conclusions, as the particle regime does not extend significantly beyond the thermosphere boundary in the line-of-sight projected views shown in Figure~\ref{fig: helium_cloud}.

\subsection{Atmospheric escape} \label{subsec: atmos_escape}

The degeneracy between the mass-loss and the temperature of the upper atmosphere can be addressed by analyzing the spectrally resolved absorption spectrum. A rigorous approach involves jointly fitting the low-resolution data presented here (NIRISS-SOSS) with previous ground-based observations \citep{allart2019_helium,kirk2020_helium,spake2021_helium}. However, unlike previous studies, we detected pre-transit absorption. Upon examining the observational window of previous high-resolution data, we found that the out-of-transit spectrum was constructed from observations taken during the pre-transit linear slope phase we observe (see Figure~\ref{fig: helium_trans_lc}). Ground-based detection of such subtle absorption is challenging, especially considering the long exposures required to match our S/N ratio. This raises the possibility of biases in high-resolution ground-based observations, potentially affecting both the shape and level of the measured absorption spectrum. Another important consideration is the potential time variability of the signal. The chemistry of metastable helium is strongly influenced by the stellar XUV flux, which is known to vary over time due to stellar activity, magnetic cycles, and other factors. A correlation between XUV variability and absorption measurements has been observed in other systems (e.g., WASP-69\,b; \citet{exo_aeronomy_wasp69b}), though not to the same extent. The previous high-resolution observations of WASP-107b, while mutually consistent, span just over a year, making it difficult to rule out time variability. In contrast, the precision of the \textit{JWST} light curve reveals details that may be missed by high-resolution spectroscopy, reinforcing the case for coordinated high-resolution follow-up during an extended \textit{JWST} baseline. Consequently, we chose not to use the high-resolution data to constrain the shape of the absorption spectrum generated by \texttt{EvE}. We verified that the amplitude of our observed signal is consistent with with the previously observed ground-based magnitudes.

As previously mentioned, our analysis suffers from a degeneracy between mass-loss rate and temperature. Additionally, the H/He ratio, a crucial parameter influencing upper atmospheric chemistry and absorption signal shape, remains uncertain. Due to limitations in the available high-resolution data, we fixed this ratio to the solar value of 90:10, a common practice in the literature. Moreover, the super-solar metallicity we derive in Section \ref{subsec: atmos_retrieve} suggests that more advanced chemistry is needed to accurately model the thermosphere. For instance, \citet{Linssen2024high_metallicities} showed that the mass-loss rate can vary by up to a factor of 3 depending on the metallicity. As a result, we cannot draw definitive conclusions about the upper atmospheric density structure. Our estimated mass-loss rate and temperature are indicative and may not accurately reflect the planet's true parameters. To resolve these parameter degeneracies, high-resolution observations with a longer baseline are necessary.

The high sensitivity of the NIRISS-SOSS light curve provides a strong constraint on the shape of the upper atmosphere. We modeled the pre- and post-transit ellipsoidal distortions separately, as illustrated in Figure~\ref{fig: helium_cloud}. The observed pre-transit light curve is well-reproduced by ellipses with sizes ranging from 10 to 18 planetary radii, which project to 8.5 to 15 planetary radii along the line-of-sight (see Figure~\ref{fig: helium_cloud}). While we varied the ellipse size along other axes, we found no clear preference, as this parameter is degenerate with the density profile and, consequently, mass-loss rate and temperature. Our post-transit absorption analysis is limited by the observational window and NIRISS-SOSS's resolution, which prevents us from assessing the trailing tail's dynamics. As a result, we cannot precisely constrain the post-transit outflow shape or determine whether it is in a particle or fluid regime. However, we estimate that the absorption duration will exceed previous ground-based reports \citep{spake2021_helium}, emphasizing the need for complementary ground-based and space-based observations. Notably, \cite{MacLeod2024} predicted a hybrid bubble-stream structure for WASP-107\,b's atmosphere through hydrodynamic simulations. The upper atmosphere can be seen as an ellipsoidal dense stream along the orbit that remains close to the planet. This prediction aligns with the shape used in our simplified modeling approach.

\subsection{Atmospheric retrievals} \label{subsec: atmos_retrieve}

We performed free atmospheric retrievals on the (R\,$\sim$\,100) transmission spectrum of WASP-107\,b, extracted from \texttt{exoTEDRF}, to constrain its atmospheric composition and vertical temperature-pressure profile. We removed the wavelengths and flux corresponding to the helium lines when running the retrievals. We employed four retrieval frameworks: \texttt{SCARLET}, \texttt{petitRADTRANS} (pRT), \texttt{Pyrat Bay}, and \texttt{TauREx}. All four retrievals used free chemistry in their analyses. We included the following species in the retrievals: Na, K, H$_2$O, CO, CO$_2$, CH$_4$, and NH$_3$ (referred to as our `definite' list of species). We also tested additional models that included additional species such as SO$_2$, PH$_3$, H$_2$S, HCN, TiO and VO (together they are called `non-definite' list). However, these species are not well constrained due to the lack of strong spectral features in the NIRISS-SOSS wavelength range and the low temperature of the planet.

\texttt{Pyrat Bay} and \texttt{pRT} included a gray cloud deck with a retrieved patchiness fraction to account for partial cloud coverage. \texttt{TauREx}, on the other hand, included a solid gray cloud deck. \texttt{Pyrat Bay} and \texttt{pRT} also implemented a short-wavelength Rayleigh-style scattering slope. In contrast, the Rayleigh-style cloud model in \texttt{TauREx} is species-specific \citep{taurex_rayleigh}, which did not fit the data well. Additionally, \texttt{TauREx} used low-resolution opacities. As a result, we do not provide any interpretation of the results from \texttt{TauREx}. However, we include the retrievals in the figures for completeness. Our analysis suggests that stellar contamination is the most likely cause of the blueward slope in the spectrum. As \texttt{SCARLET} can robustly account for stellar surface heterogeneities in addition to planetary atmospheric properties, we selected it as our primary framework and present its results here. We also used the best-fitting SOSS order 2 offset (137.67 ppm) from \texttt{SCARLET} retrievals to analyze SOSS data within the other frameworks.


\subsubsection{SCARLET}


We performed free chemistry retrievals to interpret the \textit{JWST} NIRISS-SOSS spectrum of WASP-107\,b using the \texttt{SCARLET} atmospheric retrieval framework \citep{benneke_distinguishing_2012,benneke_characterizing_2013,benneke_strict_2015,benneke_sub-neptune_2019,benneke_water_2019,pelletier_where_2021,piaulet_evidence_2023,Piaulet_Ghorayeb_2024}. This approach allowed us to test assumptions not supported by other codes. The \texttt{SCARLET} framework consists of a forward modeling component, which computes atmospheric properties given a set of input parameters, and a retrieval component, which efficiently samples large parameter spaces. For the chemistry, we opted to perform free chemistry retrievals (assuming constant abundance profiles across all atmospheric layers) rather than enforcing chemical consistency. This decision was motivated by findings of methane depletion, indicative of disequilibrium chemistry, on WASP-107\,b \citep{kreidberg2018_water_hst,sing2024_nirspec,wellbanks2024_nircam}.

Given the planet's T-P profile and chemical abundances (see below), \texttt{SCARLET} performs a hydrostatic equilibrium calculation and solves the radiative transfer equation to model the transmission spectrum. For each combination of model parameters, we use \texttt{scipy.minimize} to determine the best-fitting photospheric radius of the planet that matches the observed spectrum, iterating over the hydrostatic equilibrium and radiative transfer steps. This method was found to yield atmospheric inferences that are perfectly compatible in terms of precision and accuracy with other retrieval frameworks that marginalize over the planet radius (e.g. \citealp{Piaulet_Ghorayeb_2024,ahrer_escaping_2025}).

While we later test the impact of assumptions about the chemistry, temperature profile, stellar heterogeneity, and aerosol properties on our results (see below), we first describe the setup for our nominal \texttt{SCARLET} retrieval. We fit for the abundances of Na, K, H$_2$O, CO, CO$_2$, CH$_4$, and NH$_3$, using HELIOS-K \citep{grimm_helios-k_2015} to compile cross-sections from various sources (see Table \ref{tab:opacity_sources}). H$_2$ and He are treated as background filler gases, adopting a Jupiter-like ratio of He/H$_2$ = 0.157 \citep{vonZahn1996}.

\begin{table}[h!]
    \centering
    \begin{tabular}{ll}
        \toprule
        \textbf{Molecule} & \textbf{Reference} \\
        \midrule
        H$_2$O & \citet{ExoMol_H2O} \\
        CO & \citet{Hargreaves2019} \\
        CO$_2$ & \citet{ExoMol_CO2} \\
        CH$_4$ & \citet{HargreavesEtal2020apjsHitempCH4} \\
        HCN & \citet{Harris_HCN_2006} \\
        H$_2$S & \citet{ExoMol_H2S} \\
        SO$_2$ & \citet{ExoMol_SO2} \\
        NH$_3$ & \citet{ExoMol_NH3} \\
        \bottomrule
    \end{tabular}
    \caption{List of the references for the cross-sections used to compute the opacities of the molecules included in \texttt{SCARLET} retrievals.}
    \label{tab:opacity_sources}
\end{table}

We adopt a parametric form for the temperature profile \citep{MadhusudhanSeager2009apjRetrieval}. For aerosols, we fit for an enhanced scattering slope caused by small-particle hazes and a gray (wavelength-independent) cloud deck that truncates atmospheric signatures below the fitted cloud top pressure.

We also account for differences between the spectrum of the portion of the stellar disk being occulted by the planet and that of the rest of the stellar surface (which contributes to the out-of-transit stellar spectrum). These differences can introduce wavelength-dependent signatures in exoplanetary transmission spectra, known as the transit light source (TLS) effect \citep{Rackham2018ThePlanets}. Such signatures include slopes at short wavelengths (mimicking a haze scattering slope due to unocculted cooler stellar spots) and even molecular signatures, particularly for late-type hosts where water in cool stellar spots can create spurious water bands in the transmission spectrum. These effects can bias atmospheric retrieval inferences if unaccounted for \citep{iyer_influence_2020,SAG23_effect_TLSE}. Our detection of a strong short-wavelength slope in the spectrum thus motivates the inclusion of stellar spots in the nominal model. 

\begin{figure*}[]
    \centering
    \includegraphics[width=0.98\linewidth]{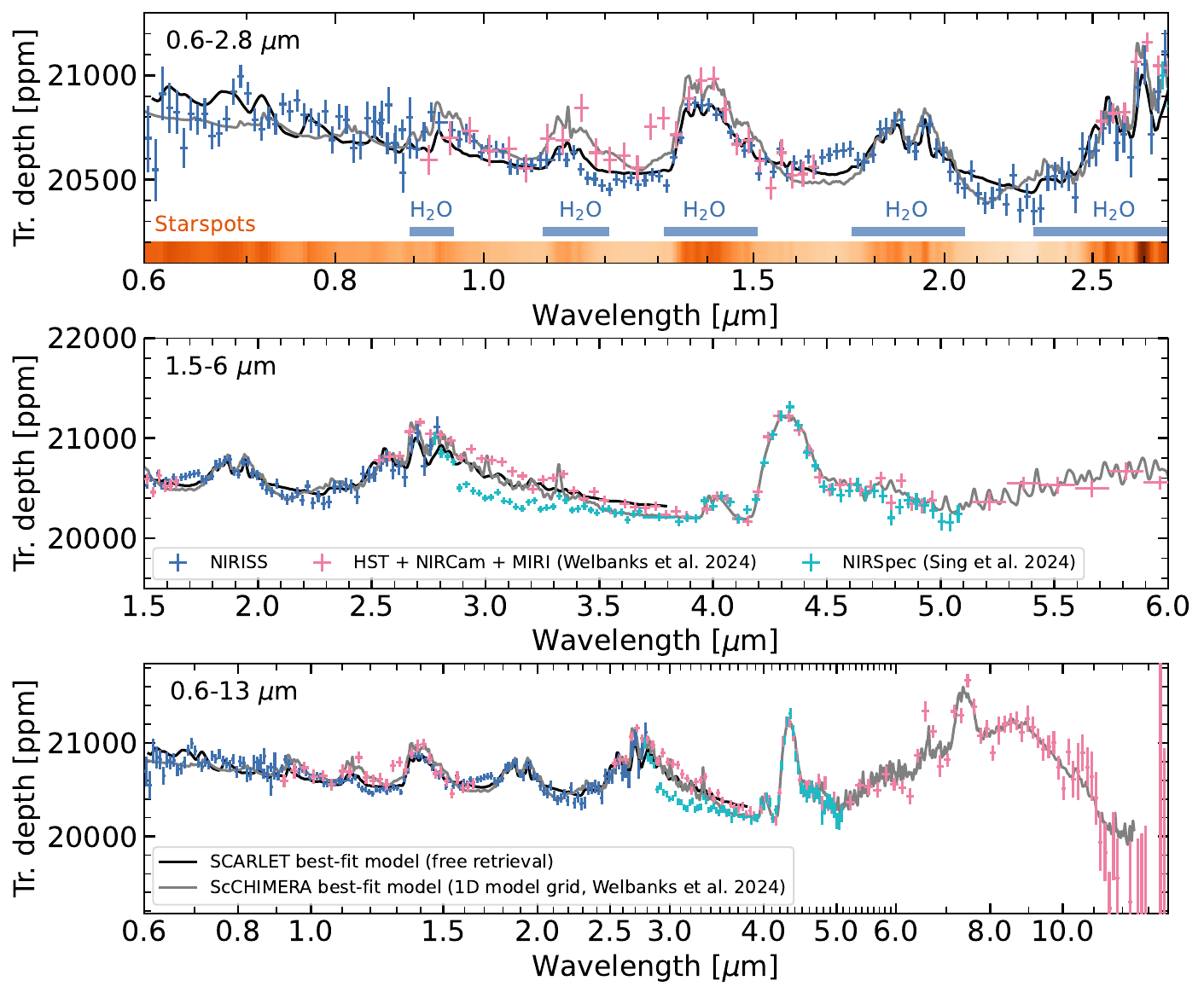}
    \caption{NIRISS-SOSS spectrum in the context of all the previous observations of WASP-107\,b. \textit{Top panel:} Transmission spectrum of WASP-107\,b over the 0.6 to 2.8$\mu$m wavelength range. The SOSS order 2 spectrum was shifted by the best-fitting offset retrieved with \texttt{SCARLET}. The NIRISS observations are shown in dark blue, while the spectrum from \citet{welbanks_high_2024} is superimposed (no offset) in pink, and the observations published in \citet{sing_warm_2024} are shown in cyan. The best-fit model from the \texttt{SCARLET} free retrieval is displayed in black, overlayed with the best-fitting model from the 1D self-consistent ScCHIMERA model grid published in \citet{welbanks_high_2024} (gray). Our NIRISS-SOSS observations reveal a strong contribution from stellar spots (shades of orange corresponding to the starspot contribution) as well as water opacity (highlighted in blue). Contrary fo Figure~\ref{fig:retrieval_spectrum_fit}, we do not offset the \textit{HST} spectrum in order to reproduce the spectrum as fitted by \citet{welbanks_high_2024}.  \textit{Middle panel:} Same thing, over the 1.5 to 6$\mu$m wavelength range. Even if the longer-wavelength data beyond 2.8$\mu$m was not included in our retrievals, we accurately predict the water slope between 2.7 and 3.8$\mu$m observed with NIRCam. We binned the NIRCam and NIRSpec spectra (3 points together for NIRCam, 8 points together for NIRSpec) for easier visual comparison, and offset the NIRSpec spectrum down by 300 ppm to obtain a match with the CO$_2$ feature observed at $\sim$4.3$\mu$m with NIRCam and NIRSpec. We note that the NIRSpec and NIRCam spectra do not find the same overall CO$_2$ amplitude and water slope  between 2.7 and 3.8$\mu$m, highlighting the importance of the NIRISS-SOSS wavelength range covering multiple water bands to avoid biases when constraining the water abundance. \textit{Bottom panel:} Full 0.6 to 13$\mu$m spectrum of WASP-107\,b. The MIRI spectrum was shifted by the median retrieved instrument offset reported in \citet{welbanks_high_2024} (282 ppm). The inclusion of the constraints on the SO$_2$, CO$_2$ and cloud opacity are needed to match the features observed beyond 3.8$\mu$m, which are not included in the \texttt{SCARLET} model because of their weak contributions over the NIRISS-SOSS wavelength range.}
    \label{fig:all_datasets_spectrum}
\end{figure*}

Finally, we also fit for an offset between the order 2 spectrum and the order 1 spectrum. For parameter sampling, we use the Nested Sampling method \citep{skilling_nested_2004,skilling_nested_2006} implemented in \texttt{nestle}\footnote{\url{http://kylebarbary.com/nestle}} within \texttt{SCARLET}, with 1,000 live points. Each model is computed at a resolving power of 15,625 (or 31,250 to test the sensitivity to model resolution) and convolved within each spectral bin, assuming a uniform throughput for likelihood evaluation. The main results from this nominal run are presented in Figures~\ref{fig:retrieval_spectrum_fit} and \ref{fig:posterior_molecules_compare}, and the prior setup and posterior constraints on the atmospheric properties are provided in Table \ref{tab:SCARLET_prior_posterior}. We find that our retrieved spectral slope and water abundance provide a good match to the NIRCam observations \citep{welbanks_high_2024}, although the consideration of stellar contamination overall leads to smaller water bands over the NIRISS-SOSS wavelength range than predicted from the degenerate water opacity slope available to anchor the water abundance with NIRSpec and NIRCam (Figure~\ref{fig:all_datasets_spectrum}).

Our results are in line with the retrieval results from the \textit{HST}+NIRCam+MIRI analysis \citep{welbanks_high_2024}. While the analysis of the \textit{JWST}/NIRSpec spectrum yielded a higher water abundance \citep{sing_warm_2024}, we caution that NIRSpec/G395H generally cannot provide precise water abundance measurements in isolation as the presence of water manifests only as a slope within its wavelength range, and especially in our case because the lack of short-wavelength coverage in NIRSpec/G395H does not allow for good constraints on the contribution of potential stellar contamination features to the water feature (and stellar contamination is not considered in the NIRSpec/G395H retrievals). We also find that NIRSpec and NIRCam recover different water opacity slopes longwards of 3$\mu$m (Fig. \ref{fig:all_datasets_spectrum}), which highlights the importance of NIRISS-SOSS for reliable water abundance constraints which are not sensitive to potential instrument offsets or model degeneracies in the absence of coverage for a full water band.

\begin{figure*}[]
    \centering
    \includegraphics[width=0.98\linewidth]{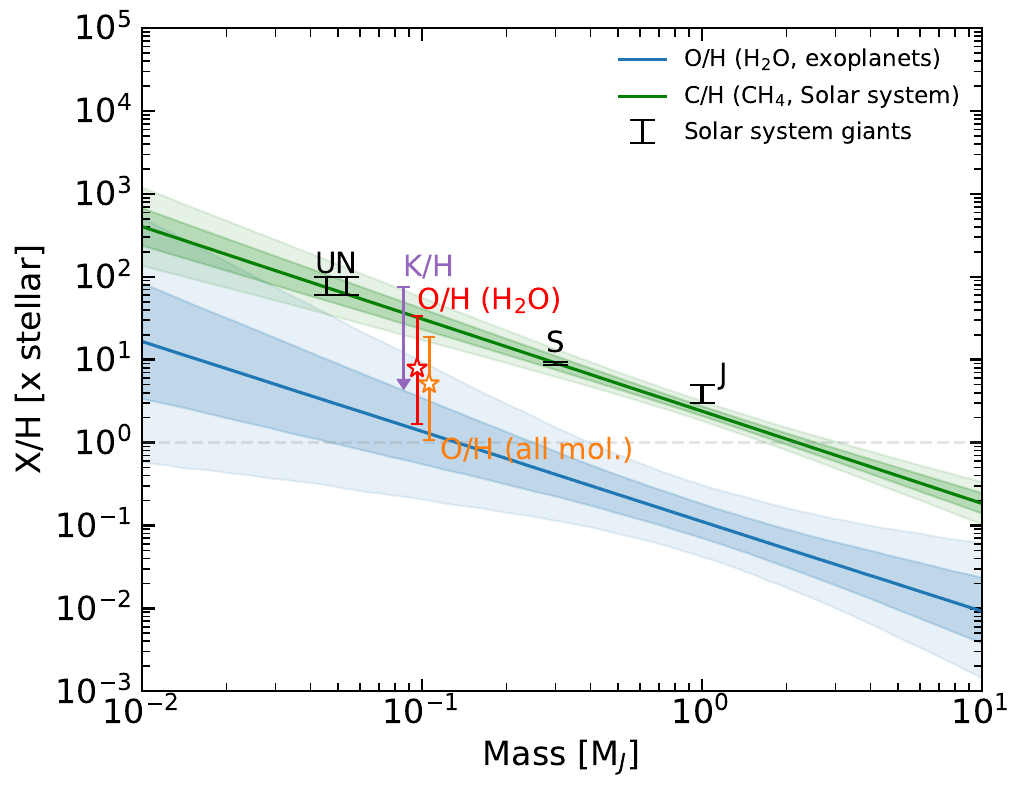}
    \caption{Composition of the atmosphere of WASP-107\,b in the context of the mass-metallicity trend. We show the median (with 1, 2$\sigma$ in semi-transparent shadings) trend in enhancement relative to solar (stellar) metallicity for the C/H ratio of the solar system giants (black points from \citet{atreya2018}, green slope) as well as the O/H ratio enhancement trend inferred for exoplanets from H$_2$O abundance measurements in gas giant atmospheres (blue trend, \citealp{welbanks_mass-metallicity_2019}). The measurement (with the \texttt{SCARLET} retrieval) for WASP-107\,b's K/H (purple, 2$\sigma$ upper limit, shifted for clarity) and O/H (red value inferred from the H$_2$O abundance alone; orange value, shifted, inferred from all the O-bearing species) is also overlaid for comparison. For WASP-107\,b, with a mass of 0.096 M$_J$, we infer a metallicity slightly higher than would be expected from the mass-metallicity trend. The upper limit we obtain on the K/H is consistent with the O/H constraints, suggesting that the atmosphere of WASP-107\,b is not significantly enhanced in refractory elements.}
    \label{fig:mass_metallicity}
\end{figure*}

We put our results in context by computing the enhancement we measure in the K/H and O/H ratios relative to what would be expected from the star's elemental abundances. As the baseline, we assume that the stellar [O/H] and [K/H] follow the measured [Fe/H] of 0.020 \citep{piaulet_wasp-107bs_2021} in their enhancements relative to solar values. We use the values from \citet{asplund_chemical_2009} for the photospheric abundances of K, O, and H of the Sun. For the K/H ratio, we obtain a 2$\sigma$ upper limit of $K/H<75\times$ stellar. Finally, we calculate two values for the O/H: one that is derived only from the H$_2$O abundances and assumes that half of the elemental oxygen is contained in H$_2$O (O/H = $8^{+26}_{-6}\times$ stellar), and a second one that is derived from the posterior distributions on all the oxygen-bearing molecules considered in the retrieval, with the caveat that major absorption bands of these species are not available within the NIRISS-SOSS bandpass (O/H= $5^{+13}_{-4}\times$ stellar). Finally, we compare these values to the expectation from the H$_2$O-derived O/H mass-metallicity trend for exoplanets \citep{welbanks_mass-metallicity_2019} and to the CH$_4$-derived C/H mass-metallicity trends of the solar system giant planets \citep{atreya2018} (see Figure~\ref{fig:mass_metallicity}).

We used Bayesian model comparison to evaluate the robustness of our detection of the TLS effect, and obtain a $\Delta \ln Z=11.62$ between a model that includes both spots and hazes, and one that only accounts for the atmosphere contribution (with hazes), which is translated to a 5.2$\sigma$ evidence for the presence of spots \citep{trotta_bayes_2008,benneke_characterizing_2013}. We additionally perform a retrieval where we add the contribution of stellar faculae (hotter than the photosphere), but this model is strongly disfavored ($\Delta \ln Z=-5.55$). Additionally, we try allowing for a partial cloud coverage by jointly fitting the fraction $f_\mathrm{cloud}$ of the terminator which is covered by clouds (as implemented in \citealp{Piaulet_Ghorayeb_2024}), but find that this addition is not warranted from a Bayesian model comparison perspective (we retrieve $f_\mathrm{cloud}=0.99 \pm 0.01$, in line with the findings of \citep{welbanks_high_2024}).

\subsubsection{petitRADTRANS}

We performed a free chemistry retrieval using \texttt{petitRADTRANS} (pRT) \citep{molliere_prt1}, fitting for planet radius, equilibrium temperature, internal temperature, cloud structure, Rayleigh scattering slope, and volumetric abundances of H$\mathrm{_2}$O, CO, CO$\mathrm{_2}$, CH$\mathrm{_4}$, Na, K, and NH$\mathrm{_3}$ (i.e., our definite list). We also tested our retrievals for additional atmospheric species such as H$\mathrm{_2}$S, HCN, PH$\mathrm{_3}$, and SO$\mathrm{_2}$ (i.e., `non-definite' list with optical absorbers like TiO and VO). However, the data still prefers a model with our definite list to the list with other additional species. Hence we report the results from our definite list. To determine the temperature profile, we used the thermal structure from \citet{guillot_tp1} with a user-defined cloud function to parameterize the power-law dependence on cloud opacity, similar to the model used in \cite{prt_clouds}. 

The cross sections for Na and K were taken from \citet{molliere_prt1}, for H$\mathrm{_2}$O from \citet{ExoMol_H2O}, for NH$\mathrm{_3}$ from \citet{ExoMol_NH3}, and for CO, CO$\mathrm{_2}$, and CH$\mathrm{_4}$ from the HITEMP database \citep{Rothman2013TheDatabase, HargreavesEtal2020apjsHitempCH4}. Our cross-sections had an initial resolution of 1,000,000, but we binned them down to 15,000 and fit the abundances of the gases mentioned above. We also used hydrogen and helium to fill the atmosphere as a background in the ratio He/H$_2$ = 0.157 \citep{kreidberg2018_water_hst}. Gas continuum opacities for the background gases (H$_2$-H$_2$ and H$_2$-He) are accounted through collision-induced absorption \citep[CIA;][]{Borysow1988,Borysow2002}. We ran the retrievals with 2,000 live points to sample the posteriors using the \texttt{MultiNest} package. 

Our best-fit parameters are shown in Table \ref{tab:pRT_prior_posterior}. We found that our \texttt{pRT} models fit best without invoking the species from the non-definite list. The lack of strong spectral bands for these species could be the reason. Our best-fit model favors an irradiation temperature of 638.82\,K and we extract an internal temperature upper-limit of 314\,K at 2\,$\sigma$. We predict that, in an attempt to capture the lower-wavelength slope and the ``dip'' at 2-2.4\,$\mu$m \citep{feinstein_early_2023}, the model overestimates the abundance of certain species (like potassium) in the atmosphere.

\subsubsection{Pyrat Bay}

We used the \texttt{Pyrat Bay} package (v1.1.6) to perform a free chemistry retrieval of the transmission spectrum of WASP-107\,b \citep{PyratBay}. \texttt{Pyrat Bay} utilizes the open-source Bayesian statistics package \texttt{MC3} \citep{mc3} to sample the posterior parameter space using the Snooker Differential-Evolution Markov Chain algorithm \citep{terBraak2008}. In our retrievals, we fit for temperature, planet radius, molecular abundances, the planet's reference pressure, Rayleigh opacity parameters, an opaque gray cloud deck, and a patchy cloud parameter. Collision-induced absorption (CIA) by H$_2$-H$_2$ and H$_2$-He pairs is accounted for using the Borysow models \citep{Borysow1988,Borysow2002}. For the temperature parameters, we applied the \citet{guillot_tp1} thermal structure under the model of \citet{Line2013}.

We fit for the abundances of H$\mathrm{_2}$O, CO, CO$\mathrm{_2}$, CH$\mathrm{_4}$, and NH$\mathrm{_3}$, which use cross-section opacities from ExoMol (H$\mathrm{_2}$O,  CH$\mathrm{_4}$, and NH$\mathrm{_3}$) and HITRAN (CO, CO$\mathrm{_2}$) at a resolution of R=15,000. The abundances of Na and K were derived using models of sodium and potassium resonant lines from \citet{Burrows2000}. We set a cap on the threshold for cumulative trace abundances at 50\% but the retrieved median He/H$_2$ ratio = 0.164. Rayleigh opacity parameters were fit using the \citet{Lecavelier2008} Rayleigh model, which modifies absorption strength as a function of wavelength: $\kappa(\lambda) = f\kappa_0(\frac{\lambda}{\lambda_0})^\alpha$, where $\lambda_0 = 0.35\mu m$, $\kappa_0 = 5.31$  $\mathsf{x}$ $10^{-27} cm^2$. The parameters $\log(f)$ and $\alpha$ were fit in the retrieval, with default values of $\log(f) = 0.0$ and $\alpha = -4$ representing the Rayleigh opacity of H$_2$. Our priors and retrieved parameters are shown in Table \ref{tab:PyratBay_prior_posterior}. \texttt{Pyrat Bay} was able to retrieve a high internal temperature (T$_{\rm int,med}$ = 208.7\,K, with a 1$\sigma$ upper limit of 393.79\,K)  from the NIRISS-SOSS dataset. This is still lower than the estimates from NIRSpec+MIRI (460\,K; \citep{sing2024_nirspec}) and \textit{HST}+NIRCam+MIRI ($>$\,345\,K; \citep{wellbanks2024_nircam}), which had access to refractories from MIRI wavelengths. 


\subsubsection{TauREx}

The \texttt{TauREx}\,3.0 forward model \citep{Taurex} employs the temperature-pressure profile outlined in equation 49 of~\citet{guillot_tp1}. The retrieval code uses a modified two-stream approximation from equation 19 in~\citet{Line_2012}, with the fitting parameters defined as T$_{\rm irr}$ (the planet’s equilibrium temperature), k$_{\rm irr}$ (the mean infrared opacity), k$_{\rm v1}$ (the mean optical opacity one), k$_{\rm v2}$ (the mean optical opacity two), and $\alpha$ (the ratio between k$_{\rm v1}$ and k$_{\rm v2}$). The atmosphere is divided into 100 uniformly spaced layers on a logarithmic grid, ranging from $10^{5}$ to $10^{-1}$ Pa. It is composed of H$_2$ and He, with a 0.176 fill ratio, and includes the following spectroscopically active gases: CO$_2$, H$_2$O, CH$_4$, CO, NH$_3$, K, and Na.

The molecular cross sections are sourced from ExoTransmit~\citep{ExoTransmit}, which have a resolution of 1k. There are no ktables that are used in this retrieval code. The forward model accounts for absorption, collision-induced absorption (CIA), and Rayleigh scattering. HITRAN~\citep{Hitran} CIA data are used for various molecule-molecule interactions, specifically H$_2$-H$_2$ and H$_2$-He. A uniform abundance profile is assumed for all molecules with respect to the height of the planet’s atmosphere. Additionally, a simple cloud model is implemented, introducing an infinitely absorbing cloud deck into the atmosphere.

Stellar parameters for WASP-107, such as temperature, radius, and mass, as well as the planetary mass and radius of WASP-107\,b, are taken from~\citet{wasp107b_caroline2021}. The retrieval is performed using the Nestle Optimizer package, with the mixing ratios of the active gases, planetary radius, cloud deck altitude, and temperature-pressure profile variables treated as free parameters.

Rayleigh scattering is not an independent parameter that can be controlled in the \texttt{TauREx} code. Our best-fit parameters are shown in the Table \ref{tab:TauREx_prior_posterior}. Consequently, we believe that \texttt{TauREx} estimates higher abundances of CO, H$_2$O, and K compared to other retrieval codes to fit the slope at shorter wavelengths. When we attempted to constrain the abundances of CO, H$_2$O, and K in the priors, the retrievals instead estimated higher abundances of other species to fit the slope. As a result, we do not make any inferences from the results of the \texttt{TauREx} retrieval, and plot it as a dashed line in Figure~\ref{fig:retrieval_spectrum_fit}. The retrieved values and their corresponding posterior distribution as shown in Table~\ref{tab:TauREx_prior_posterior} and Figure~\ref{fig: corner taurex}.

\subsection{Retrieval stabilities} \label{subsec: retrieval_stability}

Beyond the nominal runs, we explored the impact of our assumptions regarding the molecular composition -- particularly how these assumptions influence the atmospheric mean molecular weight -- by performing retrievals with and without the chemical species from our ``non-definite'' list (see Section \ref{subsec: atmos_retrieve} for details).  Retrievals using only our definite list of species were consistently preferred across all retrieval frameworks and are therefore the ones presented in our results. The corresponding inferred parameters are shown in Tables \ref{tab:SCARLET_prior_posterior}, \ref{tab:pRT_prior_posterior}, \ref{tab:PyratBay_prior_posterior}, and \ref{tab:TauREx_prior_posterior}, with the respective posterior distributions provided in Appendix Section \ref{append: retieval corners}.

We also assessed the sensitivity of our results to the T-P profile parameterization by performing an identical retrieval where we assumed a constant temperature across the entire atmosphere. Additionally, we investigated potential instrumental limitations by excluding the ``dip'' observed in the spectrum over the 2-2.4\,$\mu$m region (which is lower than model predictions in some NIRISS-SOSS datasets \citep{feinstein_early_2023} and may be of instrumental origin) and exploring the impact of assuming a non-zero offset between order 1 and order 2 in the spectrum. Furthermore, we ran a \texttt{SCARLET} retrieval accounting only for the atmospheric contribution to the spectrum, without including a stellar spot component.

\begin{figure*}[]
    \centering
    \includegraphics[width=0.9\linewidth]{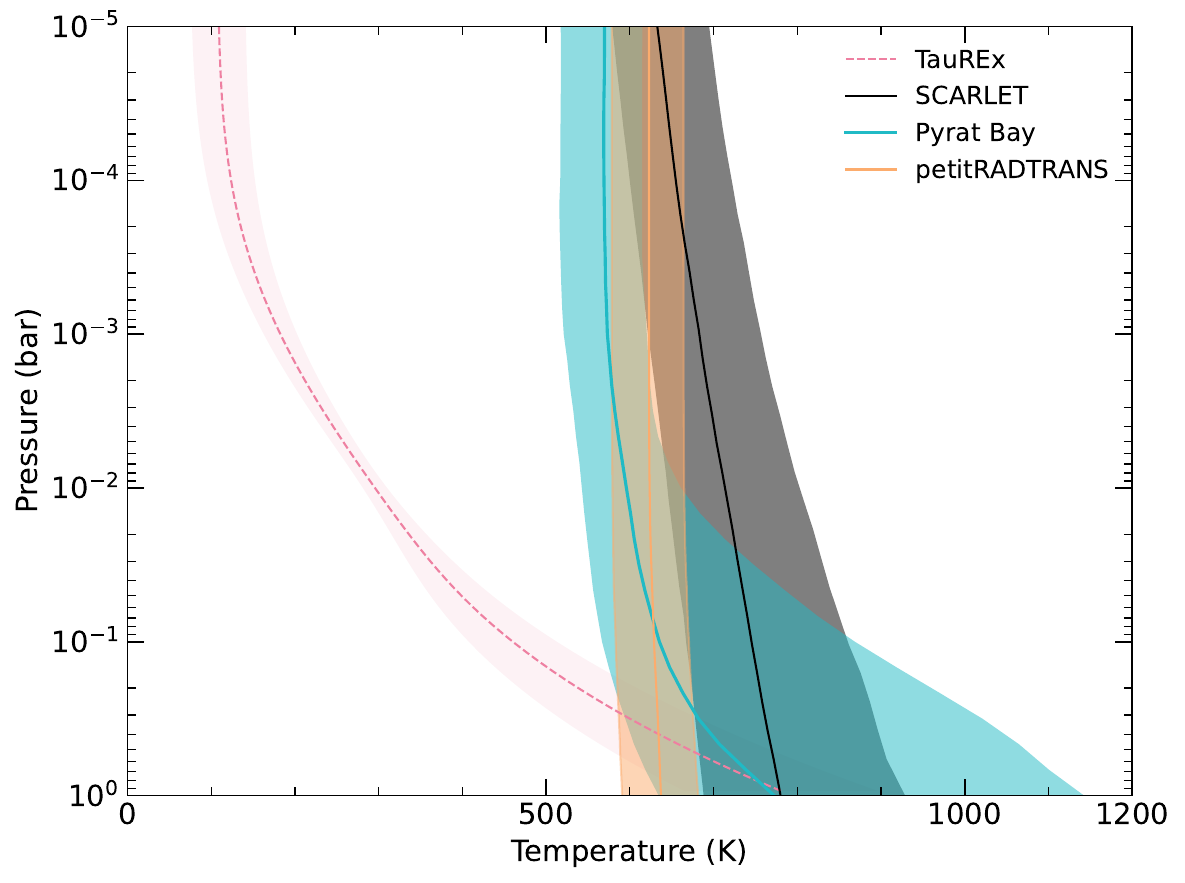}
    \caption{Temperature-pressure profiles for the \texttt{SCARLET} (nominal setup), \texttt{TauREx}, \texttt{Pyrat Bay} and \texttt{petitRADTRANS} (`definite' retrieval model) retrievals. The colors match Figure~\ref{fig:retrieval_spectrum_fit}. For each retrieval, we show the median, 1 sigma contours from the posterior samples on the T-P profile. The \texttt{SCARLET}, \texttt{Pyrat Bay} and \texttt{petitRADTRANS} models extend over a wider pressure range than the \texttt{TauREx} retrieval, and we compare them here over the pressure range where the models overlap. Although we use a non-uniform T-P profile with \texttt{SCARLET}, we retrieve a near-uniform temperature structure.}
    \label{fig:T_P_profiles}
\end{figure*}

The results are shown in Figure~\ref{fig:sensitivity_abundances}. Our H$_2$O abundance estimate is robust as long as stellar contamination is accounted for; otherwise, the retrieved water abundance is approximately 40 times higher. We also find that the retrieved temperature profile is nearly isothermal (see Figure~\ref{fig:T_P_profiles}), consistent with the agreement between the nominal retrieval and the retrieval assuming an isothermal T–P structure. \texttt{TauREx} yields varied thermal structures with unrealistically low photospheric temperatures, likely due to its inability to capture the short-wavelength slope and the dip at 2-2.4\,$\mu$m. As a result, we do not use \texttt{TauREx} to interpret any findings. \texttt{Pyrat Bay} and \texttt{petitRADTRANS} (pRT) are the only retrieval frameworks in our setups capable of constraining the internal temperature. \texttt{Pyrat Bay} provides an upper limit of $<$\,562\,K, while \texttt{pRT} constrains it to an upper limit of $<$\,315\,K at 2$\sigma$. The estimates from \texttt{Pyrat Bay} and \texttt{pRT} are comparable to the constraints from NIRSpec+MIRI \citep{sing2024_nirspec} and \textit{HST}+NIRCam+MIRI \citep{wellbanks2024_nircam}.

\subsection{C/O ratio and metallicity estimates: Implications on planet formation} \label{sebsec: metal_co}

Our 2$\sigma$ upper-limit on C/O $<$ 0.83 is consistent with results from NIRCam \citep{wellbanks2024_nircam} and NIRSpec \citep{sing2024_nirspec}. The absence of strong opacities from carbon-bearing species and the unconstrained CH$_4$ abundance may explain this. Based on our retrieved methane abundances, we estimate that methane is severely depleted compared to predictions from the Solar System's mass-metallicity relation  (log$_{10}$(CH$_4$/H) = -7.865, versus the predicted log$_{10}$(CH$_4$/H) = +1.509. This is consistent with previous estimates from both \textit{JWST} and \textit{HST}.

To estimate the metallicity from \texttt{SCARLET}'s water abundance, we assume that half the oxygen in the atmosphere is contained in water \citep[for T$_{\rm eq}<$1200\,K;][]{madhu2012_CtoO, welbanks_metalicity}. This yields an estimate of H$_2$O/H = 0.898$^{+1.415}_{-0.778}$ (i.e., approximately 8$\times$\,stellar). Our upper limit on K/H ($<$ 75\,$\times$\,stellar), combined with the water-derived metallicity, indicates a metal-enriched atmosphere. Recent simulations of warm super-Neptunes have shown that oxygen-bearing species become enriched—leading to a lower C/O ratio—and that overall metallicity increases \citep{metallicity_enrichment_super_neptunes2025}, supporting our measurements.

With the current estimate of metallicity, extrapolating back to the planet’s formation suggests a primordial metallicity of 4\,$\times$\,stellar assuming a thermospheric mass-loss rate of ($\sim$\,1--10\,M$_\oplus$/Gyr). Such a high rate of metal-rich gas accretion seems unlikely at birth for a close-in gas giant \citep{2019A&A...627A.127C}, suggesting that in-situ formation is improbable. 

The high internal temperature limits from this study, together with previous \textit{JWST} results and the observed non-zero eccentricity, suggest that the planet is tidally heated—potentially explaining its current puffy state. Since the tidal circularization timescale for WASP-107\,b is much shorter than the system’s age ($\tau \lesssim$\,2\,Myr vs. 3.4\,Gyr), an ultra-recent migration offers a plausible explanation. This implies that we may currently be observing the planet during the $\sim$2\,Myr window of tidal circularization, capturing ongoing atmospheric escape in real time.

An alternative explanation for the high internal heat is Ohmic dissipation \citep{tidal_wasp107b_2025}. This mechanism, driven by interactions between atmospheric winds and the planet’s magnetic field, can reproduce the observed internal luminosity under realistic assumptions about atmospheric circulation and magnetic field strength. In this scenario, tidal heating is not required, and constraints on the tidal quality factor are relaxed \citep[see][]{tidal_wasp107b_2025}. However, even in this case, it does not preclude planetary migration -- WASP-107\,b may still have undergone recent dynamical evolution to reach its current orbit.

\begin{table}[h!]
\centering
\renewcommand{\arraystretch}{1.3}
\setlength{\tabcolsep}{6pt}
\begin{tabular}{l c c}
\toprule
\textbf{Parameter} & \textbf{Posterior} & \textbf{Prior} \\
\midrule
\multicolumn{3}{l}{\textbf{Chemical Species}} \\
$\log_{10}(X_{\mathrm{H_2O}})$   & $-2.45^{+0.64}_{-0.67}$ & $\mathcal{U}(-10.0, 0)$ \\
$\log_{10}(X_{\mathrm{K}})$   & $<-4.89$ & $\mathcal{U}(-10.0, 0)$ \\
$\log_{10}(X_{\mathrm{NH_3}})$   & $<-5.76$ & $\mathcal{U}(-10.0, 0)$ \\
$\log_{10}(X_{\mathrm{CH_4}})$   & $<-5.14$ & $\mathcal{U}(-10.0, 0)$ \\
$\log_{10}(X_{\mathrm{CO_2}})$   & $-5.92^{+1.69}_{-2.36}$  & $\mathcal{U}(-10.0, 0)$ \\
$\log_{10}(X_{\mathrm{CO}})$     & $-5.54^{+2.56}_{-2.92}$ & $\mathcal{U}(-10.0, 0)$ \\
$\log_{10}(X_{\mathrm{Na}})$   & $<-3.13$ & $\mathcal{U}(-10.0, 0)$ \\

\midrule
\multicolumn{3}{l}
{\textbf{Temperature profile}} \\
$T_0$ (K)                        & $584.86^{+63.39}_{-51.97}$    & $\mathcal{U}(500, 1300)$ \\
$\alpha_1$ (K$^{-1/2}$)          & $1.49^{+0.33}_{-0.37}$  & $\mathcal{U}(0.2, 2.0)$ \\
$\alpha_2$ (K$^{-1/2}$)          & $1.16^{+0.51}_{-0.51}$  & $\mathcal{U}(0.2, 2.0)$ \\
$\log_{10}(P_1)$ (bar)           & $-1.18^{+2.16}_{-2.79}$ & $\mathcal{U}(-9.0, 2.0)$ \\
$\log_{10}(P_2)$ (bar)           & $-5.53^{+3.06}_{-2.24}$ & $\mathcal{U}(-9.0, 2.0)$ \\
$\log_{10}$($P_3$-$P_1$) (bar)           & $3.44^{+2.2}_{-2.17}$  & $\mathcal{U}(0,7.0)$ \\
\midrule
\multicolumn{3}{l}{\textbf{Aerosols}} \\
$\log_{10}(c_\mathrm{haze})$                   & $-4.12^{+3.41}_{-3.56}$  & $\mathcal{U}(-10, 5)$ \\
$\log_{10}(P_{\mathrm{cloud}})$ (bar) & $-3.84^{+0.7}_{-0.65}$ & $\mathcal{U}(-8.0, 2.0)$ \\
\midrule
\multicolumn{3}{l}{\textbf{Stellar contamination}} \\
$f_{\mathrm{spot}}$ (\%) & $4.74^{+0.35}_{-0.36}$ & $\mathcal{U}(0, 50)$ \\
$T_{\mathrm{phot}}$ (K) & $4409.63^{+26.79}_{-28.04}$ & $\mathcal{N}(4425,70)$ \\
$\Delta T_{\mathrm{spot}}$ (K) & $-981.58^{+25.48}_{-12.93}$ & $\mathcal{U}(-1000, -50)$ \\
\midrule
\multicolumn{3}{l}
{\textbf{Instrument offset}} \\
$\Delta_{\mathrm{O2}}$ (ppm)   & $137.67^{+21.0}_{-21.3}$    & $\mathcal{U}(-250, 250)$ \\
\bottomrule
\end{tabular}
\caption{Constraints on the atmosphere and stellar heterogeneity properties from the \texttt{SCARLET} free retrieval, along with the prior adopted on each parameter (uniform priors except on the photosphere temperature $T_\mathrm{phot}$). We report the median and $\pm 1\sigma$ range for all parameters, or 2$\sigma$ upper limits where applicable.}
\label{tab:SCARLET_prior_posterior}
\end{table}

\begin{table}[h!]
\centering
\renewcommand{\arraystretch}{1.3}
\setlength{\tabcolsep}{6pt}
\begin{tabular}{l c c}
\toprule
\textbf{Parameter} & \textbf{Best Fit} & \textbf{Prior} \\
\midrule
\multicolumn{3}{l}{\textbf{Physical Properties}} \\
R$_p$ (R$_{\rm Jup}$)  & $0.85_{-0.01}^{+0.01}$ & $\mathcal{U}$[0.8, 1.1]   \\
\midrule
\multicolumn{3}{l}
{\textbf{Thermal and \texttt{pRT} parameters}} \\
T$_\mathrm{irr}$ (K)   & $638.82_{-45.96}^{+42.27}$ & $\mathcal{U}$[400, 900.0]  \\
T$_\mathrm{int}$ (K)    & $<314.89$ & $\mathcal{U}$[100.0, 600.0] \\
$\log_{10}$($\kappa_{\rm IR}$)   & $-2.74_{-0.34}^{+0.29}$ & $\mathcal{U}$[-4.0, 2.0]\\
$\log_{10}$($\gamma_{\rm Guillot}$)  & $0.93_{-0.06}^{+0.04}$         & $\mathcal{U}$[0.5, 1.0] \\
\midrule
\multicolumn{3}{l}
{\textbf{Chemical species}} \\
$\log_{10}(X_{\mathrm{H_2O}})$  & $-2.55_{-0.96}^{+0.85}$ & $\mathcal{U}$[-12.0, -0.1]  \\
$\log_{10}(X_{\mathrm{CO}})$   & $-5.65_{-1.11}^{+0.92}$ & $\mathcal{U}$[-12.0, -0.1]      \\
$\log_{10}(X_{\mathrm{CO_2}})$  & $-3.06_{-0.42}^{+0.44}$  & $\mathcal{U}$[-12.0, -0.1]     \\
$\log_{10}(X_{\mathrm{CH_4}})$  & $<-6.13$ & $\mathcal{U}$[-12.0, -0.1]    \\
$\log_{10}(X_{\mathrm{K}})$  & $<-5.57$ & $\mathcal{U}$[-12.0, -0.1]    \\
$\log_{10}(X_{\mathrm{Na}})$  & $<-3.94$ & $\mathcal{U}$[-12.0, -0.1]    \\
$\log_{10}(X_{\mathrm{NH_3}})$  & $<-6.55$  & $\mathcal{U}$[-12.0, -0.1]     \\
\midrule
\multicolumn{3}{l}
{\textbf{Clouds and Hazes}} \\
$\log_{10}$P$_{\rm cloud}$ (bar)  & $-3.09_{-0.36}^{+0.38}$         & $\mathcal{U}$[-6.0,2.0] \\
Haze factor & $2.01_{-0.38}^{+0.36}$         & $\mathcal{U}$[-4.0,10.0] \\
$\log_{10}$($\kappa_{\rm o}$) & $7.44_{-7.83}^{+6.11}$ & $\mathcal{U}$[0.01, 20.0]\\
$\gamma_{\rm power-law}$ & $-9.32_{-3.33}^{+3.56}$  & $\mathcal{U}$[-20, 2.0]\\
Patchiness & $0.64_{-0.17}^{+0.15}$  & $\mathcal{U}$[0, 1.0]\\

\bottomrule
\end{tabular}
\caption{Atmospheric constrains from the \texttt{petitRADTRANS} free retrieval, along with the prior adopted on each parameter. We report the median and $\pm 1\sigma$ range for all parameters, or 2$\sigma$ upper limits where applicable.}
\label{tab:pRT_prior_posterior}
\end{table}

\begin{table}[h!]
\centering
\renewcommand{\arraystretch}{1.3}
\setlength{\tabcolsep}{6pt}
\begin{tabular}{l c c}
\toprule
\textbf{Parameter} & \textbf{Median Fit} & \textbf{Prior} \\
\midrule
\multicolumn{3}{l}{\textbf{Physical Properties}} \\
R$_p$ (R$_\oplus$)   & $9.63_{-0.11}^{+0.12}$ & $\mathcal{U}$[8.4, 12.7]   \\
\midrule
\multicolumn{3}{l}
{\textbf{Thermal and \texttt{Pyrat Bay} parameters}} \\
T$_\mathrm{irr}$ (K)   & $636.6_{-65.9}^{+56.9}$ & $\mathcal{U}$[500.0, 1600.0]  \\
T$_\mathrm{int}$ (K)     & $< 562.07$ & $\mathcal{U}$[100.0, 800.0] \\
log($\kappa$')   & $-5.43_{-1.01}^{+1.37}$ & $\mathcal{U}$[-7.0, 1.0]\\
log($\gamma_1$)         & $-1.21_{-0.99}^{+0.97}$     & $\mathcal{U}$[-3.0, 3.0] \\
log($\gamma_2$)          & $-1.40_{-0.89}^{+0.90}$         & $\mathcal{U}$[-3.0, 3.0] \\
$\alpha$          & $0.469_{-0.313}^{+0.368}$              & $\mathcal{U}$[0.0, 1.0]  \\
log(P$_\mathrm{ref}$) (bar) & $-3.236_{-3.776}^{+3.433}$  & $\mathcal{U}$[-9.0, 2.0] \\
\midrule
\multicolumn{3}{l}
{\textbf{Chemical species}} \\
$\log_{10}(X_{\mathrm{H_2O}})$            & $-1.868_{-0.377}^{+0.285}$ & $\mathcal{U}$[-12.0, -0.05]  \\
$\log_{10}(X_{\mathrm{CO}})$           & $< -2.1$ & $\mathcal{U}$[-12.0, -0.05]      \\
$\log_{10}(X_{\mathrm{CO_2}})$        & $-3.51^{+ 0.44}_{- 0.62}$  & $\mathcal{U}$[-12.0, -0.05]     \\
$\log_{10}(X_{\mathrm{CH_4}})$          & $< -4.36$ & $\mathcal{U}$[-12.0, -0.05]    \\
$\log_{10}(X_{\mathrm{K}})$               & $-3.94_{-1.23}^{+0.84}$ & $\mathcal{U}$[-12.0, -0.05]    \\
$\log_{10}(X_{\mathrm{Na}})$               & $<-4.20$ & $\mathcal{U}$[-12.0, -0.05]    \\
$\log_{10}(X_{\mathrm{NH_3}})$        & $<-5.63$  & $\mathcal{U}$[-12.0, -0.05]     \\
\midrule
\multicolumn{3}{l}
{\textbf{Clouds and Hazes}} \\
$\log_{10}$(f$_\mathrm{ray}$)  & $3.43_{-0.43}^{+0.33}$ & $\mathcal{U}$[0.0, 10.0]      \\
$\alpha_\mathrm{ray}$ & $-2.66_{-0.44}^{+0.55}$  & $\mathcal{U}$[-6.0, 10.0]     \\
$\log_{10}$(P$_\mathrm{top}$) (bar) & $-4.68_{-0.31}^{+0.64}$  & $\mathcal{U}$[-9.0, 2.0]     \\
f$_\mathrm{patchy}$     & $0.955_{-0.017}^{+0.016}$ & $\mathcal{U}$[0.0, 1.0]      \\
\bottomrule
\end{tabular}
\caption{Atmospheric constrains from the \texttt{Pyrat Bay} free retrieval, along with the prior adopted on each parameter. We report the best-fit and median with $\pm 1\sigma$ range for all parameters, or 2$\sigma$ upper limits where applicable..}
\label{tab:PyratBay_prior_posterior}
\end{table}

\begin{table}[h!]
\centering
\renewcommand{\arraystretch}{1.3}
\setlength{\tabcolsep}{6pt}
\begin{tabular}{l c c}
\toprule
\textbf{Parameter} & \textbf{Posterior} & \textbf{Prior} \\
\midrule
\multicolumn{3}{l}{\textbf{Physical Properties}} \\
R$_{p}$ (R$_{\rm Jup}$)               &  $0.93^{-0.06}_{+0.01}$  &   $\mathcal{U}(0.749-1.13)$  \\
\midrule
\multicolumn{3}{l}
{\textbf{Thermal and \texttt{TauREx} parameters}} \\
T$_{\rm irr}$          &  $<814.97$    & $\mathcal{U}(100-900)$  \\
$\log{\kappa_{\rm irr}}$       & $-1.30_{-1.81}^{+0.58}$  & $\mathcal{U}(-5,0)$  \\
$\log{\kappa_{\rm v1}}$       & $-2.66_{-1.48}^{+1.44}$  & $\mathcal{U}(-5,0)$   \\
$\log{\kappa_{\rm v2}}$       & $-2.66_{-1.44}^{+1.46}$  & $\mathcal{U}(-5,0)$    \\
$\log{\rm \alpha}$            &  $0.46_{-0.30}^{+0.30}$   & $\mathcal{U}(-4,0)$  \\
\midrule
\multicolumn{3}{l}
{\textbf{Chemical species}} \\

$\log_{10}(X_{\mathrm{CH_4}})$ &  $<-5.47$   & $\mathcal{U}(-12,-5)$ \\
$\log_{10}(X_{\mathrm{CO_2}})$              &  $<-1.43$   & $\mathcal{U}(-12,-0.3)$\\
$\log_{10}(X_{\mathrm{CO}})$            & $<-0.806$   & $\mathcal{U}(-12,-0.3)$ \\
$\log_{10}(X_{\mathrm{H_2O}})$    &   $<-0.32$   & $\mathcal{U}(-12,-0.3)$\\
$\log_{10}(X_{\mathrm{NH_3}})$             &  $<-5.52$    & $\mathcal{U}(-12,-5)$\\
$\log_{10}(X_{\mathrm{K}})$              &  $<-2.01$  & $\mathcal{U}(-12,-2)$  \\
$\log_{10}(X_{\mathrm{Na}})$              &  $<-2.22$  & $\mathcal{U}(-12,-2)$  \\
\midrule
\multicolumn{3}{l}
{\textbf{Clouds}} \\
$\log_{10}$(P$_{\rm clouds}$) (Pa) &   $<3.89$   & $\mathcal{U}(-1,4)$ \\  
\bottomrule
\end{tabular}
\caption{Atmospheric constrains from the \texttt{TauREx} free retrieval, along with the prior adopted on each parameter. We report the median and $\pm 1\sigma$ range or the $2\sigma$ upper bound for all parameters.}
\label{tab:TauREx_prior_posterior}
\end{table}

\begin{appendices}

\setcounter{figure}{0}
\renewcommand{\thefigure}{A\arabic{figure}}
\setcounter{table}{0}
\renewcommand{\thetable}{A\arabic{table}}

\section{Data reduction pipelines: \texttt{NAMELESS} reduction and \texttt{ExoTEP} light curve fitting} \label{data_red_appendix}

We also use the \texttt{NAMELESS} pipeline \citep{coulombe2023_nameless_wasp18b,feinstein2023_wasp39b,Coulombe2025highlyreflectivewhiteclouds} to reduce our \textit{JWST} NIRISS-SOSS observations of WASP-107\,b following a similar procedure to those presented in \citep{Benneke2024} and \citep{Coulombe2025highlyreflectivewhiteclouds}. Starting from the uncalibrated data, we go through the Stages 1 and 2 steps of the STScI \texttt{jwst} v1.12.5 pipeline \citep{bushouse_2023}. We subsequently correct for bad pixels, non-uniform background, remaining cosmic rays, and 1/$f$ noise at the integration level using a series of custom steps. Bad pixels are flagged by looking for outliers in the second derivative of the frames, as described in \citet{coulombe2023_nameless_wasp18b}, and their values are replaced using bicubic interpolation. The non-uniform background is subtracted by individually scaling the two regions of the model background provided by the STScI \footnote{\url{https://jwst-docs.stsci.edu/}} which are separated by the jump in flux near x$\sim$700. We also correct for any cosmic rays remaining after the jump detection step of the \texttt{jwst} pipeline by computing the running median in time of all individual pixels and clipping any value that is more than 4$\sigma$ away from its median. Finally, we correct for the 1/$f$ noise following the method described in \citet{Benneke2024,Coulombe2025highlyreflectivewhiteclouds}, where each column of the first and second spectral orders is scaled individually, considering only pixels within a 30-pixel distance from the center of the traces to compute the 1/$f$ noise. The light curves are then extracted from the first and second spectral orders using a box aperture with a width of 36 pixels.

\begin{figure*}[]
  \centering
  \includegraphics[width=\linewidth]{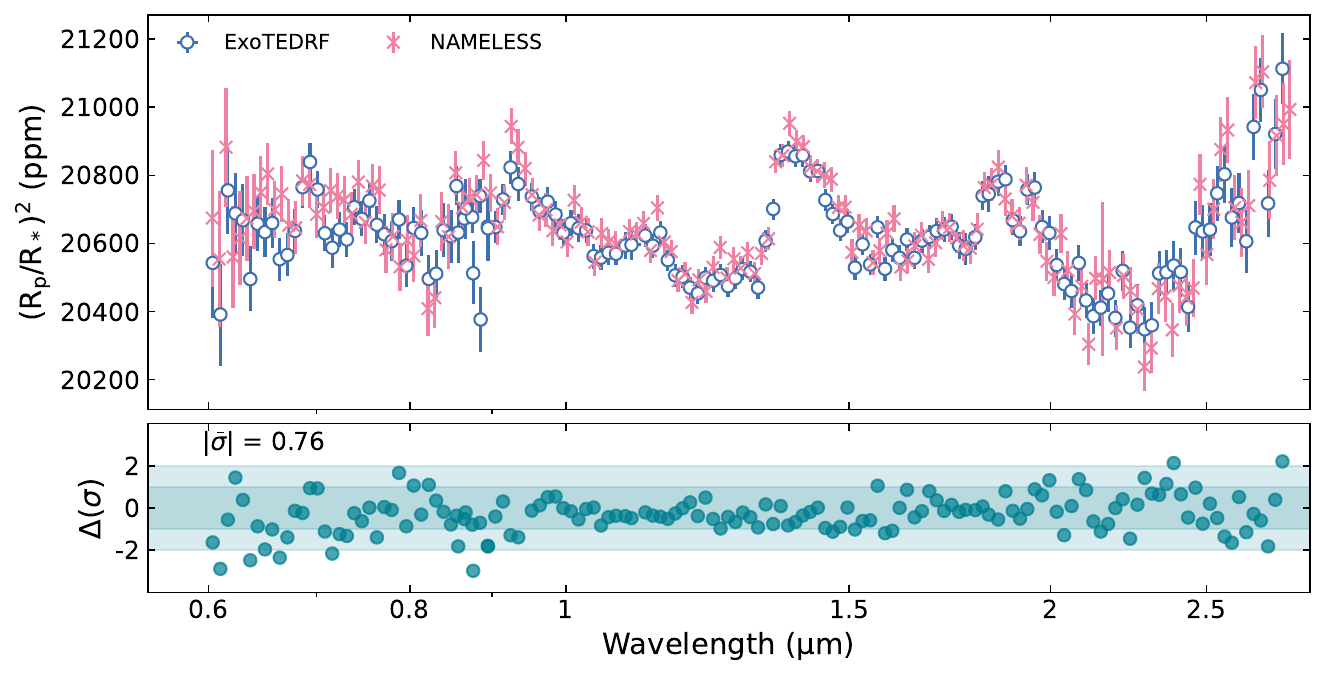}
  \caption{Transmission spectra of WASP-107\,b extracted from \texttt{NAMELESS} and \texttt{exoTEDRF} pipelines, binned down to R$\sim$100. The bottom planel shows the difference in transit depths i.e (\texttt{exoTEDRF} - \texttt{NAMELESS}) scaled by the maximum difference. There is no trend in the differences, indicating the pipelines are in good agreement with an average scatter of 0.76\,$\mathrm{\sigma}$. }
  \label{fig: comparison}
\end{figure*}

We fit the white-light and spectroscopic light curves produced by our \texttt{NAMELESS} reduction using the ExoTEP framework \citep{benneke_sub-neptune_2019,benneke_water_2019}. For the white-light curve fit, produced by summing all wavelengths of the first spectral order (0.85--2.85\,$\mu$m), we keep the time of mid-transit ($T_0$, $\mathcal{U}[2460107,2460108]$\,BJD), planet-to-star radius ratio ($R_\mathrm{p}/R_\mathrm{s}$, $\mathcal{U}$[0.01,0.2]), semi-major axis ($a/R_\mathrm{s}$, $\mathcal{U}$[10,30]), and impact parameter ($b$, $\mathcal{U}$[0,1]) as free parameters. We consider the quadratic limb-darkening law and fit directly for $u_1$ and $u_2$, assuming large uniform priors ($U$[-3,3]) to avoid the introduction of potential biases in the transmission spectrum \citep{Coulombe2024LDbiases}. For the systematics model, we consider a linear trend with slope $v$ ($\mathcal{U}$[-10$^9$,10$^9$]) and intercept $c$ ($\mathcal{U}$[-10$^9$,10$^9$]). ExoTEP models the transit light curves using the \texttt{batman} python package \citep{batman} and explores the parameter space using the Markov chain Monte Carlo (MCMC) sampler \texttt{emcee} \citep{Foreman_Mackey_2013}. We run \texttt{emcee} for 10,000 steps, discarding the first 6,000 steps as burn-in. We repeat the same procedure for the spectroscopic light curves, which we fit at the native resolution (one light curve per detector column), having fixed the orbital parameters of WASP-107\,b to the best-fit values from the white-light curve ($T_0$ = 2460107.5058791\,BJD, $a/Rs$ = 18.082, $b$ = 0.087). The resulting spectrum in shown in Figure~\ref{fig: comparison}, in comparison with the spectra from \texttt{exoTEDRF} pipeline.
We find that the overall shape of the two spectra is similar, with a mean absolute deviation of 0.76\,$\mathrm{\sigma}$. However, there is an increasing scatter between the two reductions at wavelengths below $\sim$0.85\,$\mu$m (corresponding to the order 2 data) and above $\sim$2.2\,$\mu$m, where the throughput of the instrument is significantly lower and the spectra are most sensitive to data reduction choices. These differences are most likely attributable to distinct treatments of the 1/$f$ noise and subtraction of the background, which impact the precision of the transmission spectra and dilute its amplitude. 

\begin{figure*}[]
  \centering
  \includegraphics[width=\linewidth]{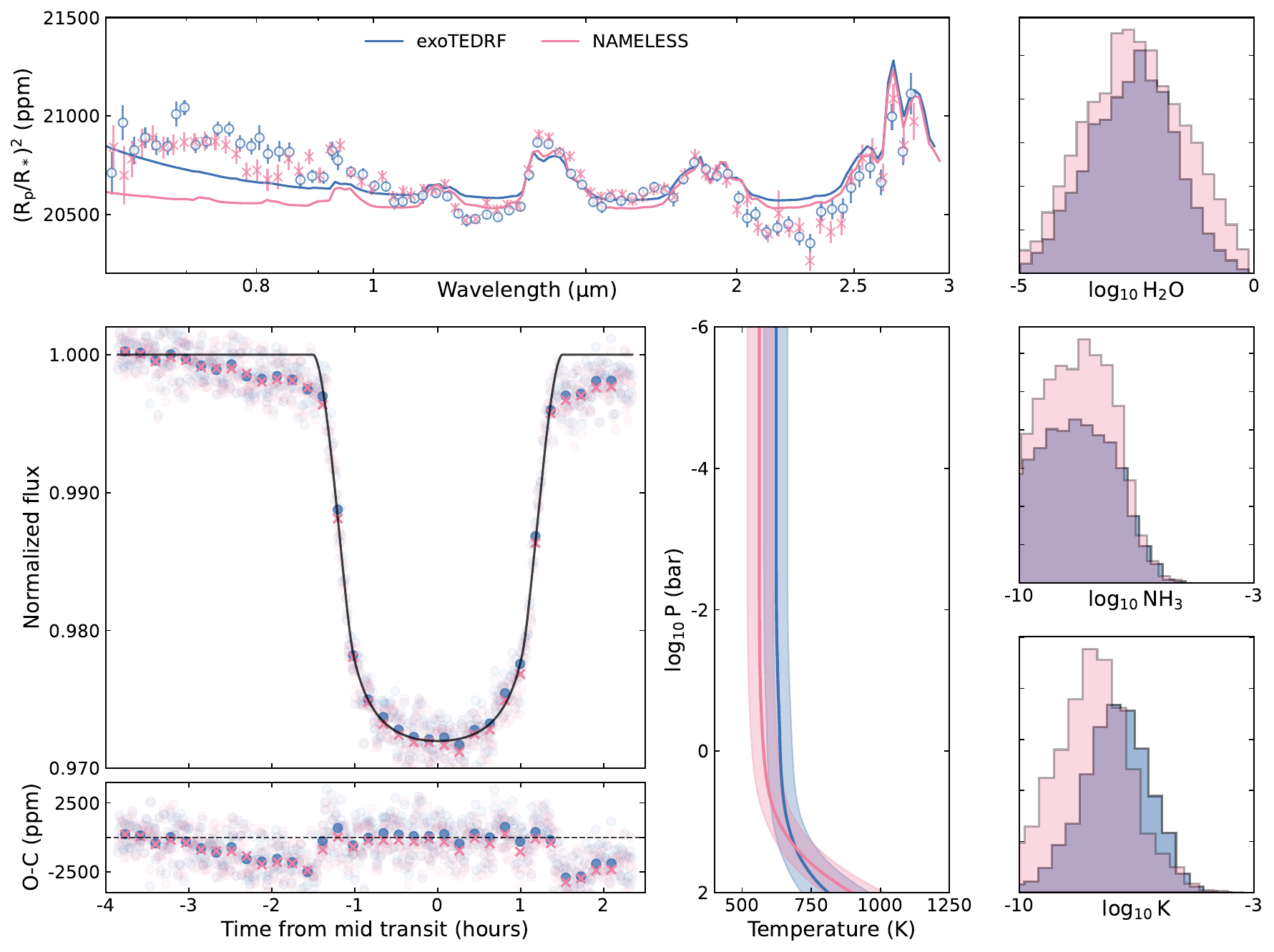}
  \caption{Free retrieval comparison of the \textit{JWST}/NIRISS-SOSS transmission spectrum from the \texttt{exoTEDRF} and \texttt{NAMELESS} pipelines. The top panel shows the \texttt{pRT} fits for both datasets. Note that both datasets were run with the best-fit offset from \texttt{SCARLET}. The \texttt{NAMELESS} pipeline produces a shallower lower-wavelength slope than \texttt{exoTEDRF}, but the extracted thermal profiles (middle column) and volumetric abundances (right column) are consistent with each other. The helium light curves (pixels corresponding to $\lambda$: 1.0830216 -- 1.0839538\,$\mu$m) from both pipelines (left panels) are also consistent.}
  \label{fig: nameless_retrieval}
\end{figure*}

To test the impact of the differences between the two spectra on the inferred atmospheric properties, we ran \texttt{pRT} retrievals on the \texttt{NAMELESS}-reduced dataset (Extended Data Fig. \ref{fig: nameless_retrieval}). We find that the measured water and potassium abundances from the two reductions are consistent at less than 1\,$\sigma$, and that they provide virtually the same NH$_3$ abundance upper limit. Moreover, we show that the extracted helium transit light curves from \texttt{NAMELESS} and \texttt{exoTEDRF} exhibit the same strong signal of pre- and post-transit absorption. 

\section{Significance of light curve fits} \label{lc_fits_append}

\begin{figure*}[]
  \centering
  \includegraphics[width=\linewidth]{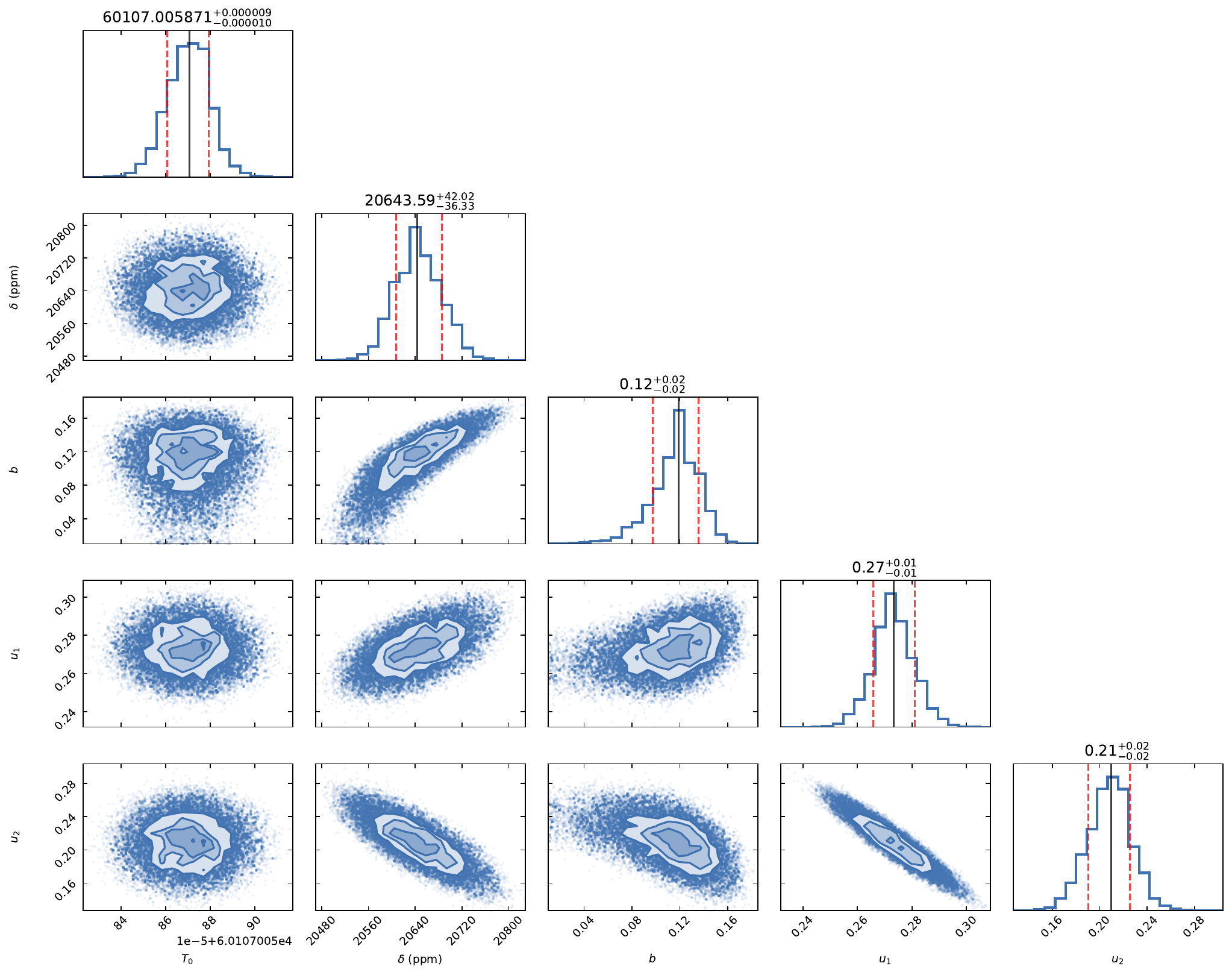}
  \caption{Corner plots for NIRISS-SOSS Order 1 broadband fits using \texttt{juliet} without any detrending performed. The solid black line shows the 0.5 quantile with the upper and lower 1$\sigma$ errors shown in red.}
  \label{fig: corner O1_no_trend}
\end{figure*}

\begin{figure*}[]
  \centering
  \includegraphics[width=\linewidth]{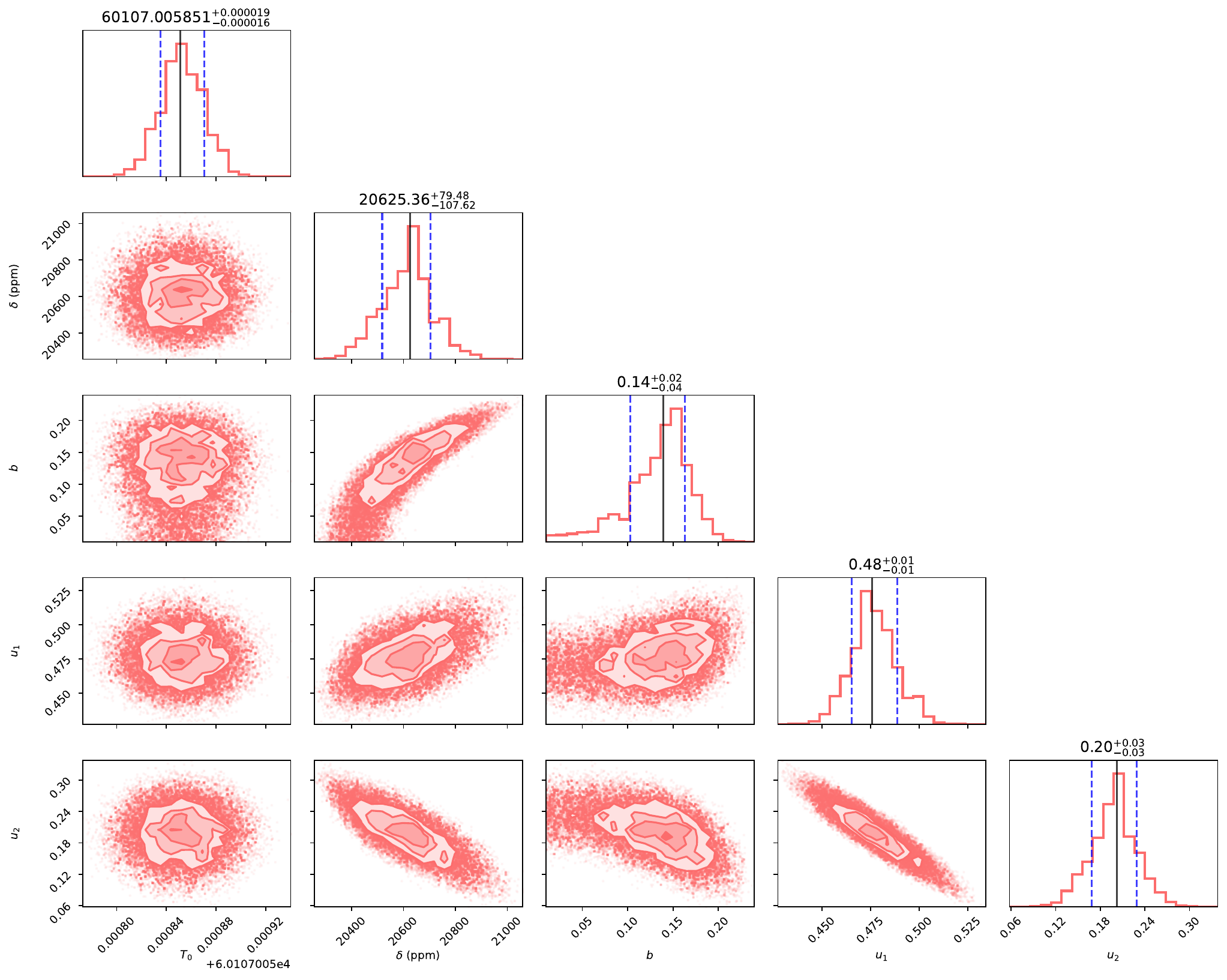}
  \caption{Corner plots for NIRISS-SOSS Order 2 broadband fits using \texttt{juliet} with linear detrending performed. The solid black line shows the 0.5 quantile with the upper and lower 1$\sigma$ errors shown in blue.}
  \label{fig: corner O2_lin_trend}
\end{figure*}

\begin{figure*}[]
  \centering
  \includegraphics[width=\linewidth]{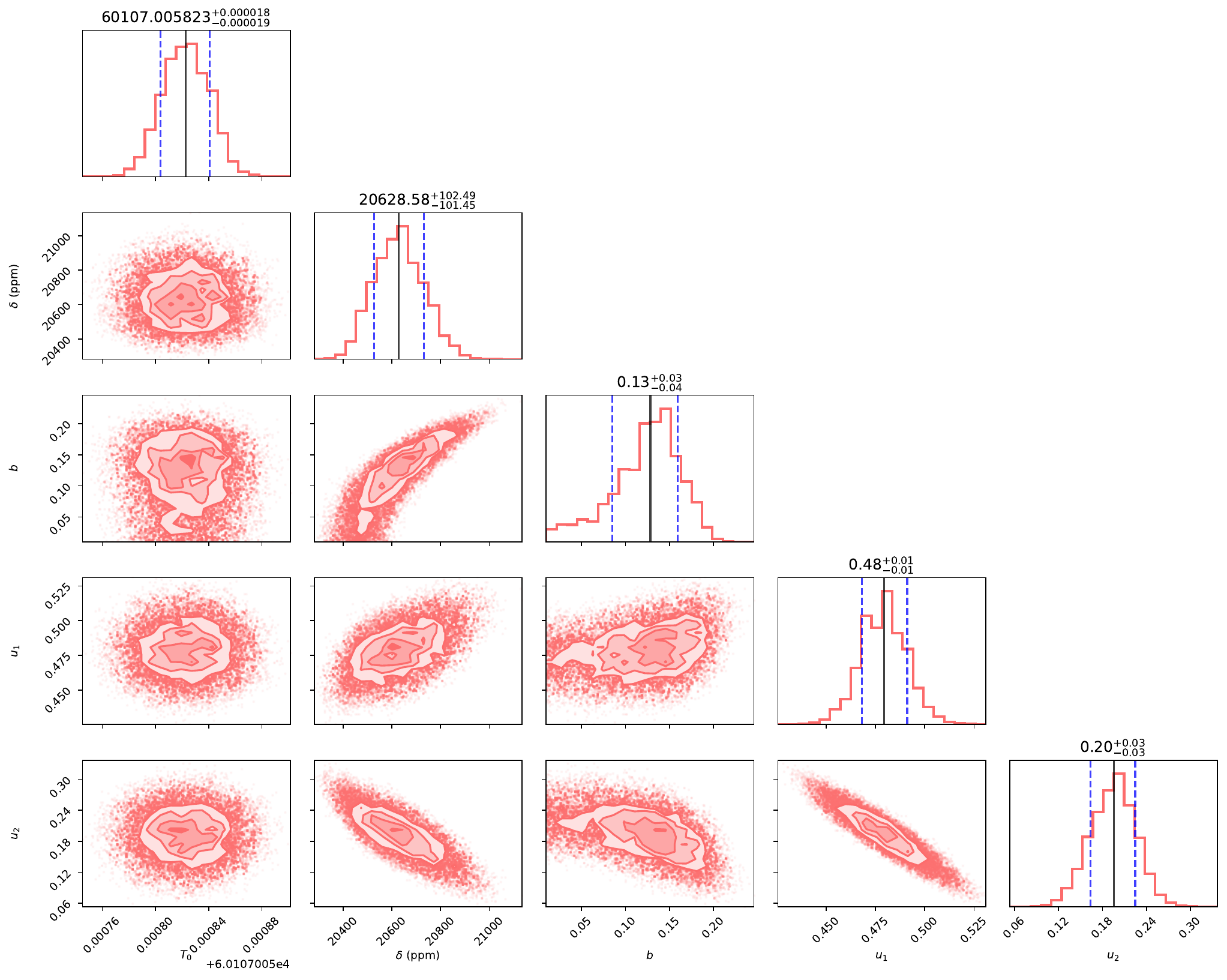}
  \caption{Corner plots for NIRISS -OSS Order 2 broadband fits using \texttt{juliet} without linear detrending performed. The solid black line shows the 0.5 quantile with the upper and lower 1$\sigma$ errors shown in blue. The BIC indicates that this model is preferred over linear detrending model.}
  \label{fig: corner O2_no_trend}
\end{figure*}

We performed linear de-trending while fitting the broadband order 2 light curve. However, the Bayesian Information Criterion (BIC) favors the model without linear de-trending, with a $\Delta$BIC of 24.936. The corner plots for order 1 without de-trending and for order 2 with and without de-trending are shown in Figures \ref{fig: corner O1_no_trend}, \ref{fig: corner O2_lin_trend}, and \ref{fig: corner O2_no_trend}, respectively. It is clear that the transit depths extracted from order 2 with and without de-trending are virtually identical, differing by only $\sim$\,3\,ppm (30$\times$ smaller than the standard deviation), and thus do not impact the nature of our results. This is particularly important in the context of pixel-resolution fits, where we aim to detect an extended helium atmosphere that begins well before and persists long after the optical transit duration. Therefore, we did not apply any de-trending algorithms.

\begin{figure*}[]
  \centering
  \includegraphics[width=\linewidth]{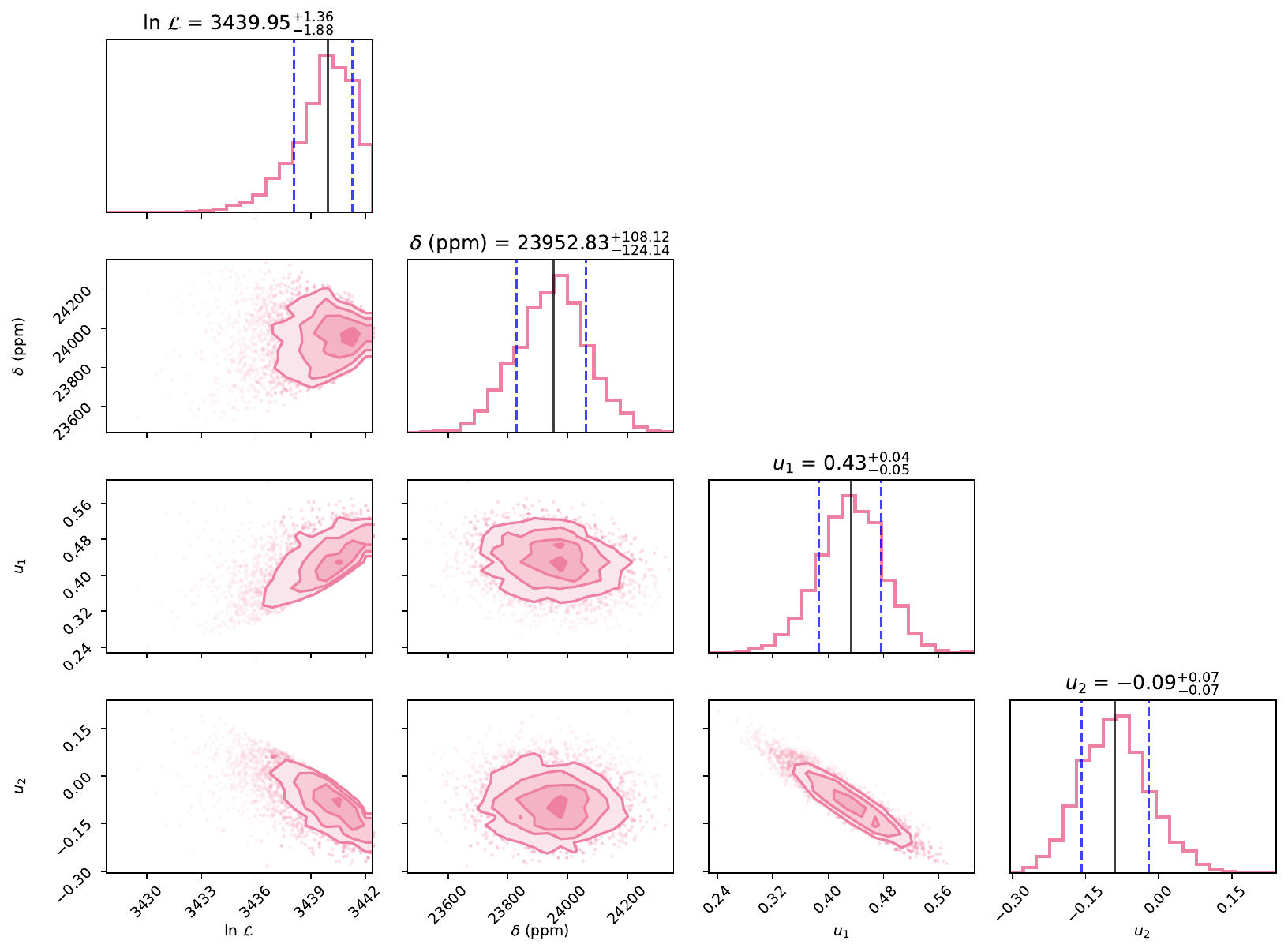}
  \caption{Corner plots for NIRISS-SOSS pixel-resolution fit for $\lambda$: 1.0835\,$\mu$m using \texttt{juliet}. The solid black line shows the 0.5 quantile with the upper and lower 1$\sigma$ errors shown in blue. }
  \label{fig: corner he_in}
\end{figure*}

\begin{figure*}[]
  \centering
  \includegraphics[width=\linewidth]{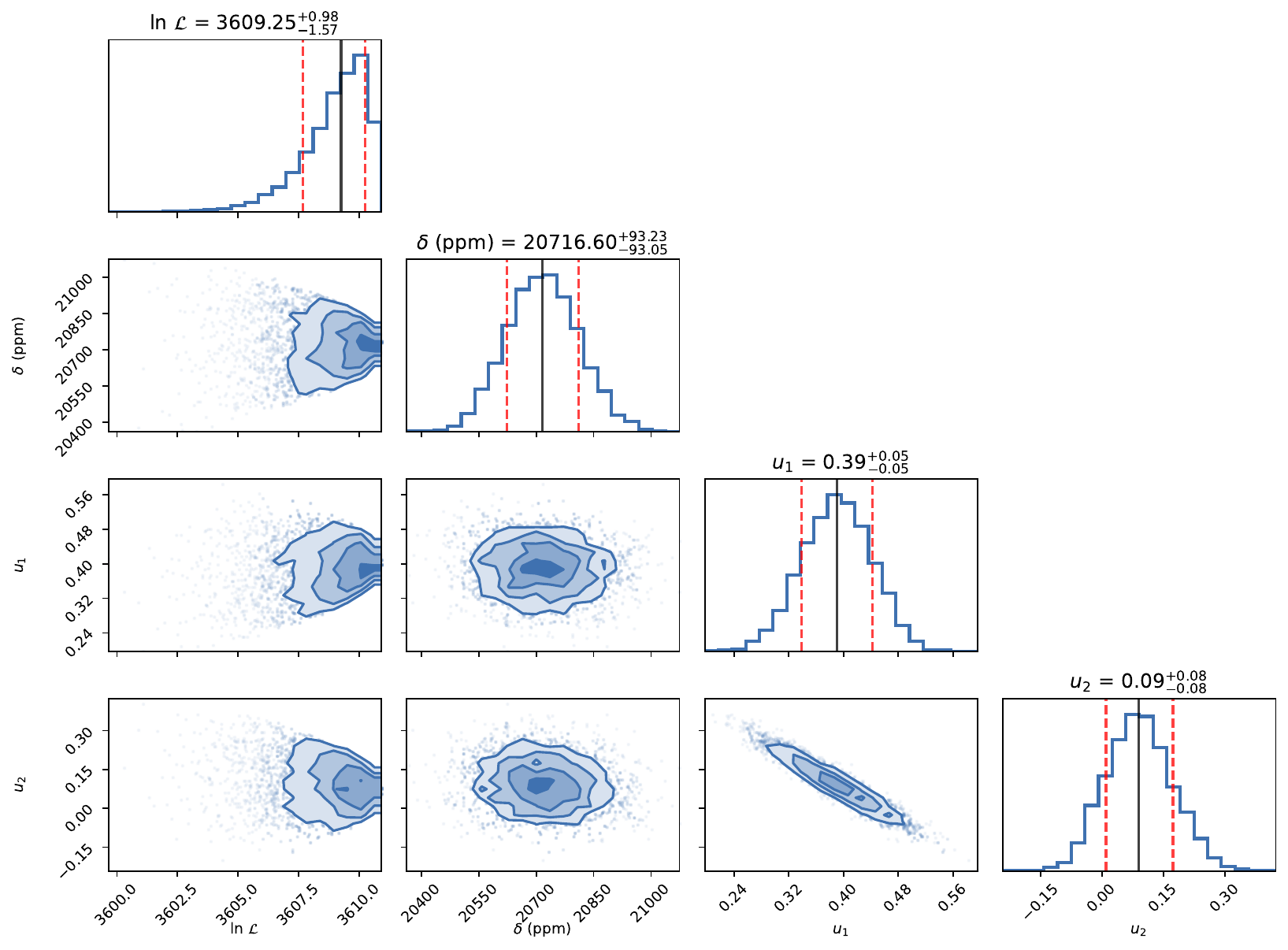}
  \caption{Corner plots for NIRISS-SOSS pixel-resolution fit for $\lambda$: 1.0835\,$\mu$m using \texttt{juliet}. The solid black line shows the 0.5 quantile with the upper and lower 1$\sigma$ errors shown in blue. }
  \label{fig: corner he_out}
\end{figure*}

The pixel-resolution light curve fit centered on the helium line ($\lambda$: 1.0835\,$\mu$m) is shown in Figure \ref{fig: corner he_in}. We can see the depth of the helium line is well-constrained and shows deeper transit depth than the one outside the line (see Figure \ref{fig: corner he_out}).

\section{Retrieval frameworks} \label{append: retieval corners}

\begin{figure*}[]
  \centering
  \includegraphics[width=\linewidth]{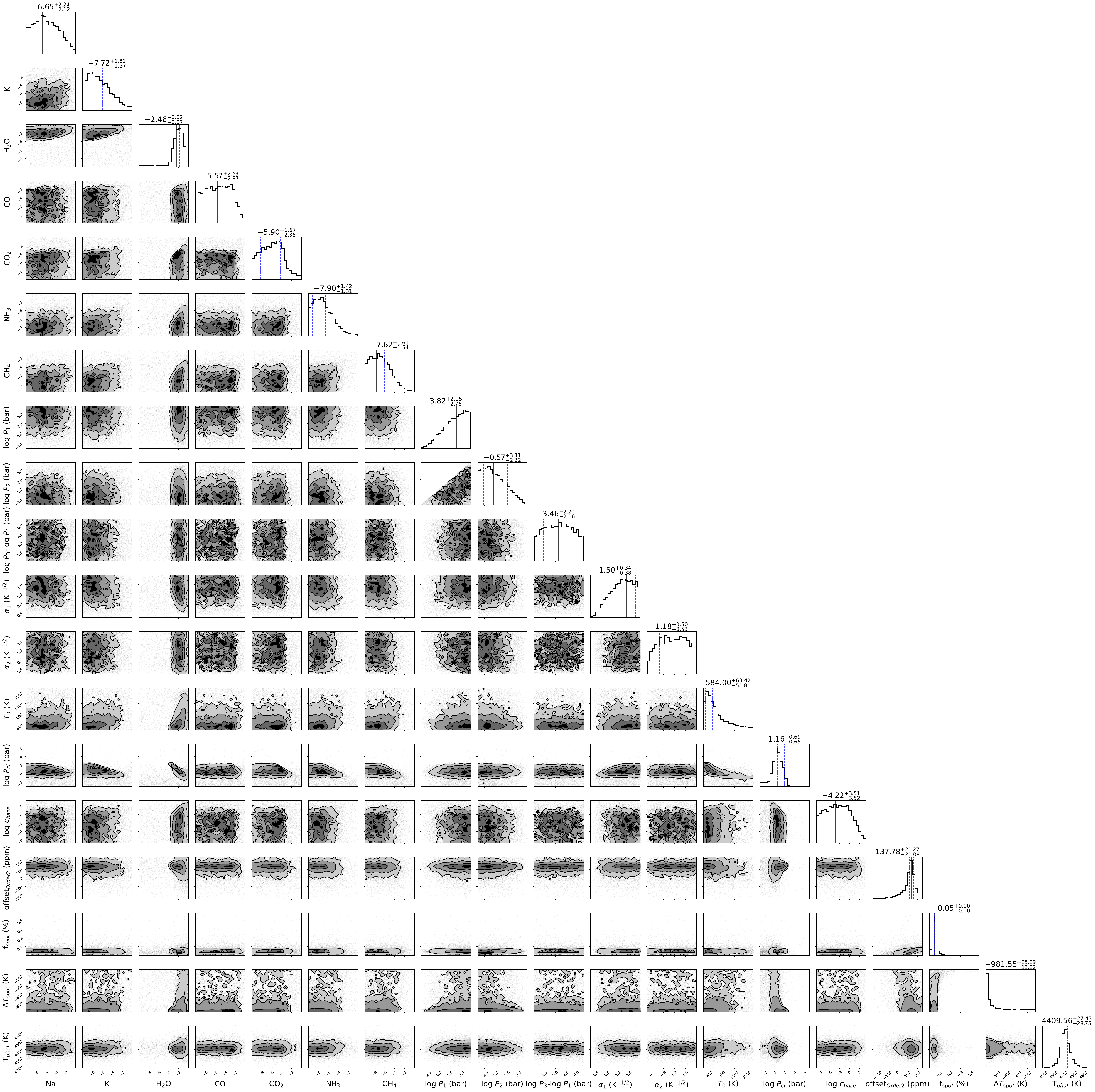}
  \caption{Corner plots from the \texttt{SCARLET} free-chemistry retrievals of the NIRISS-SOSS dataset at R$\sim$100. Median values with 1$\sigma$ uncertainties are highlighted. Volume mixing ratios of the chemical species are shown in log$_{10}$ scale. }
  \label{fig: corner scarlet}
\end{figure*}

\begin{figure*}[]
  \centering
  \includegraphics[width=\linewidth]{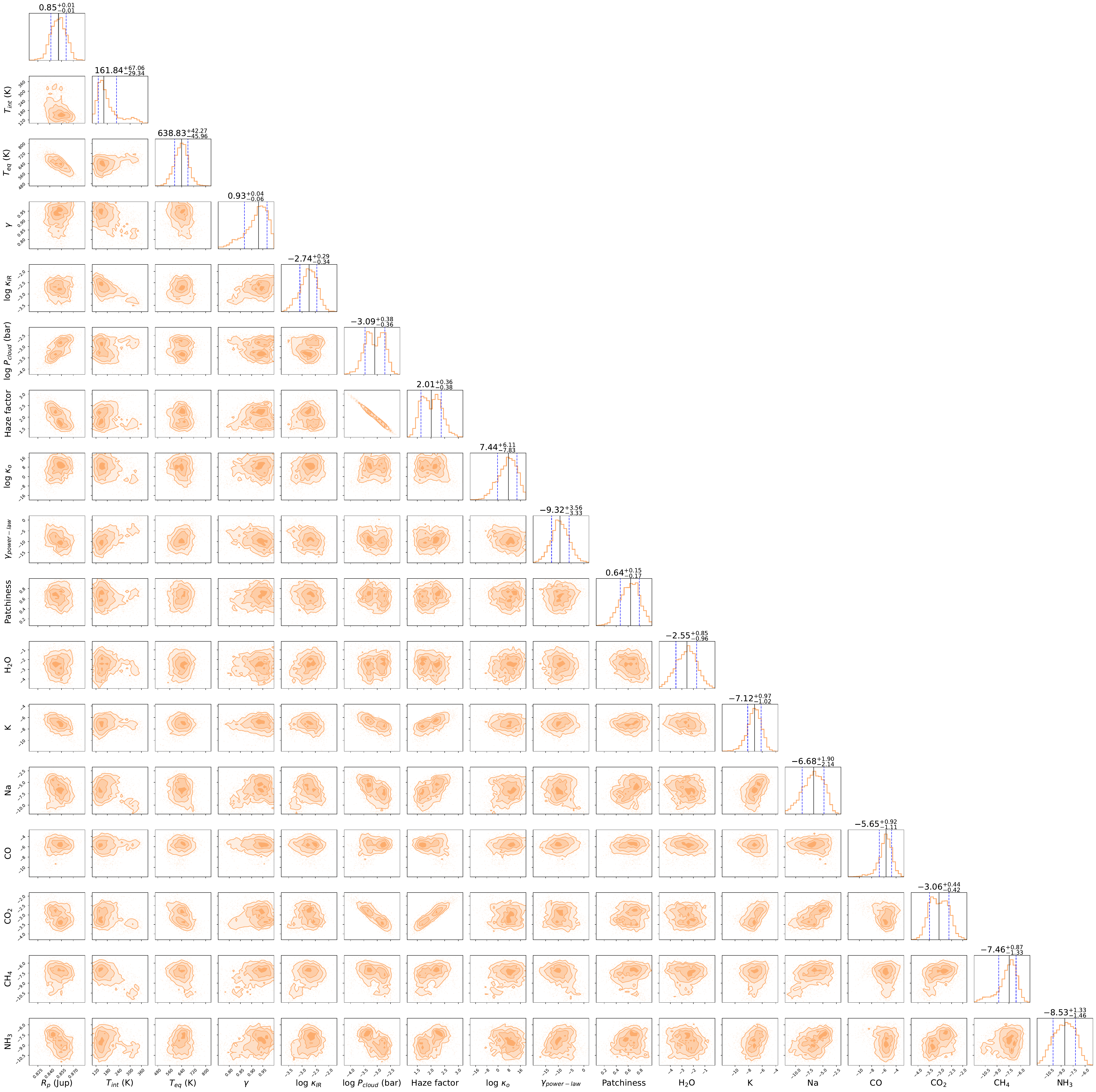}
  \caption{Same as Figure \ref{fig: corner scarlet}, but with \texttt{petitRADTRANS}.}
  \label{fig: corner prt}
\end{figure*}

\begin{figure*}[]
  \centering
  \includegraphics[width=\linewidth]{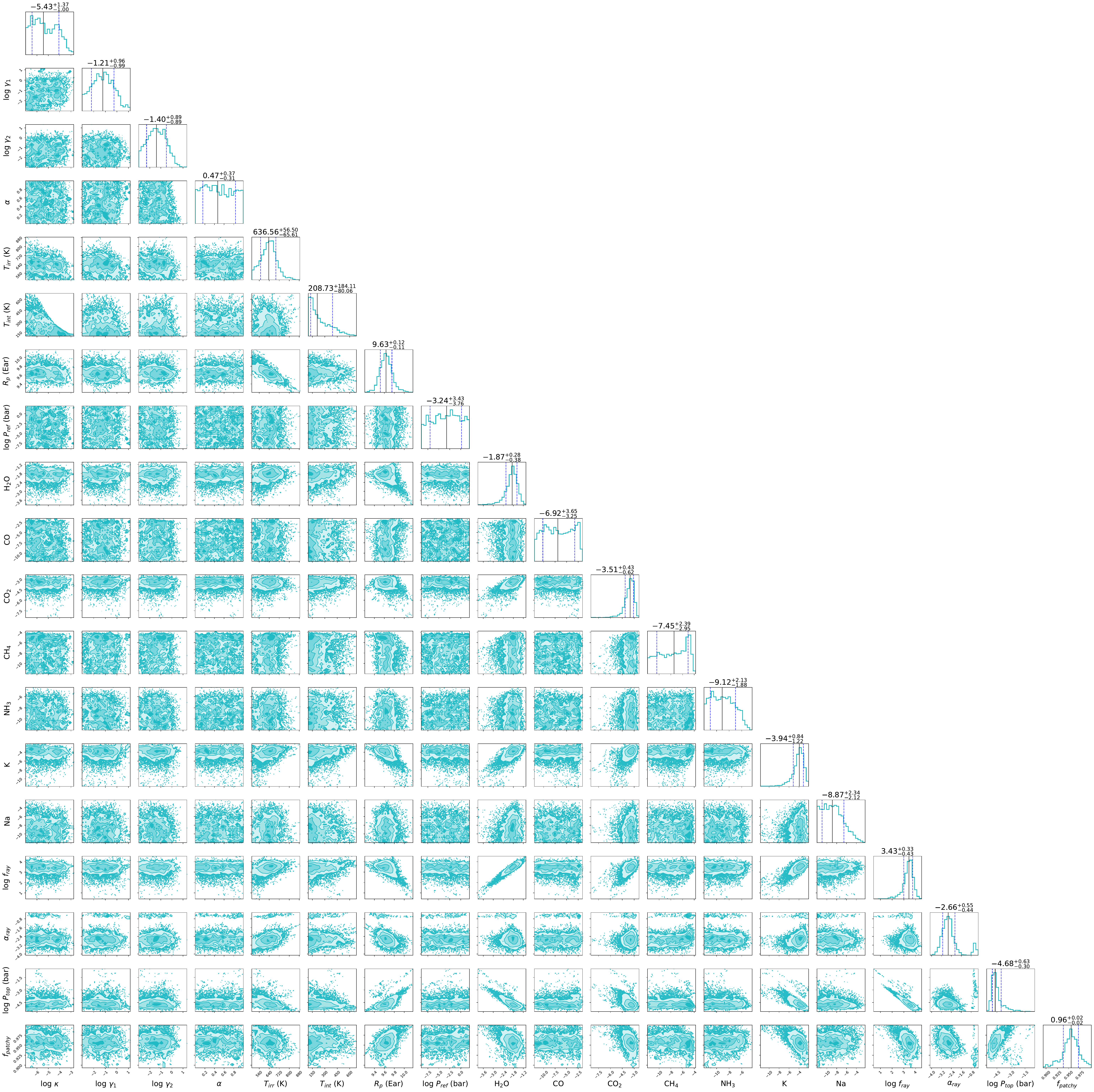}
  \caption{Same as Figure \ref{fig: corner scarlet}, but with \texttt{Pyrat Bay}.}
  \label{fig: corner pyrat}
\end{figure*}

Our retrieval frameworks -- \texttt{SCARLET}, \texttt{petitRADTRANS}, and \texttt{Pyrat Bay}—fit the overall trend of the SOSS data, with additional deviations at shorter wavelengths ($\lambda<$1$\mu$m). Since we attribute the short-wavelength slope to stellar heterogeneity rather than hazes, the \texttt{pRT} and \texttt{Pyrat Bay} retrievals are unable to capture this feature. The corner plots of the best-fit free-chemistry models from \texttt{SCARLET}, \texttt{pRT}, and \texttt{Pyrat Bay} are shown in Figures  \ref{fig: corner scarlet}, \ref{fig: corner prt}, and \ref{fig: corner pyrat}, respectively. 

\begin{figure*}[]
  \centering
  \includegraphics[width=\linewidth]{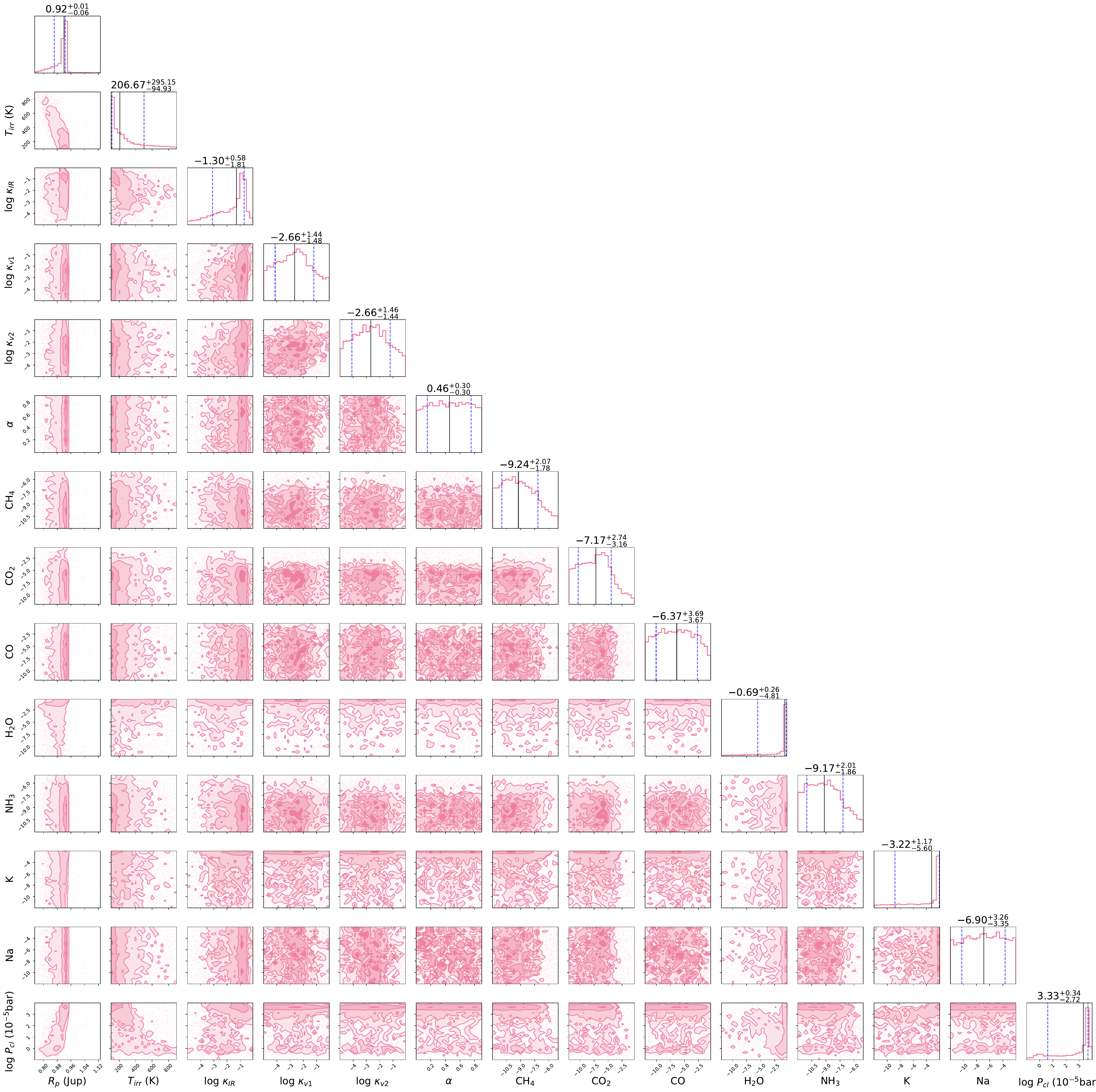}
  \caption{Same as Figure \ref{fig: corner scarlet}, but with \texttt{TauREx}.}
  \label{fig: corner taurex}
\end{figure*}

\begin{figure*}[]
  \centering
  \includegraphics[width=\linewidth]{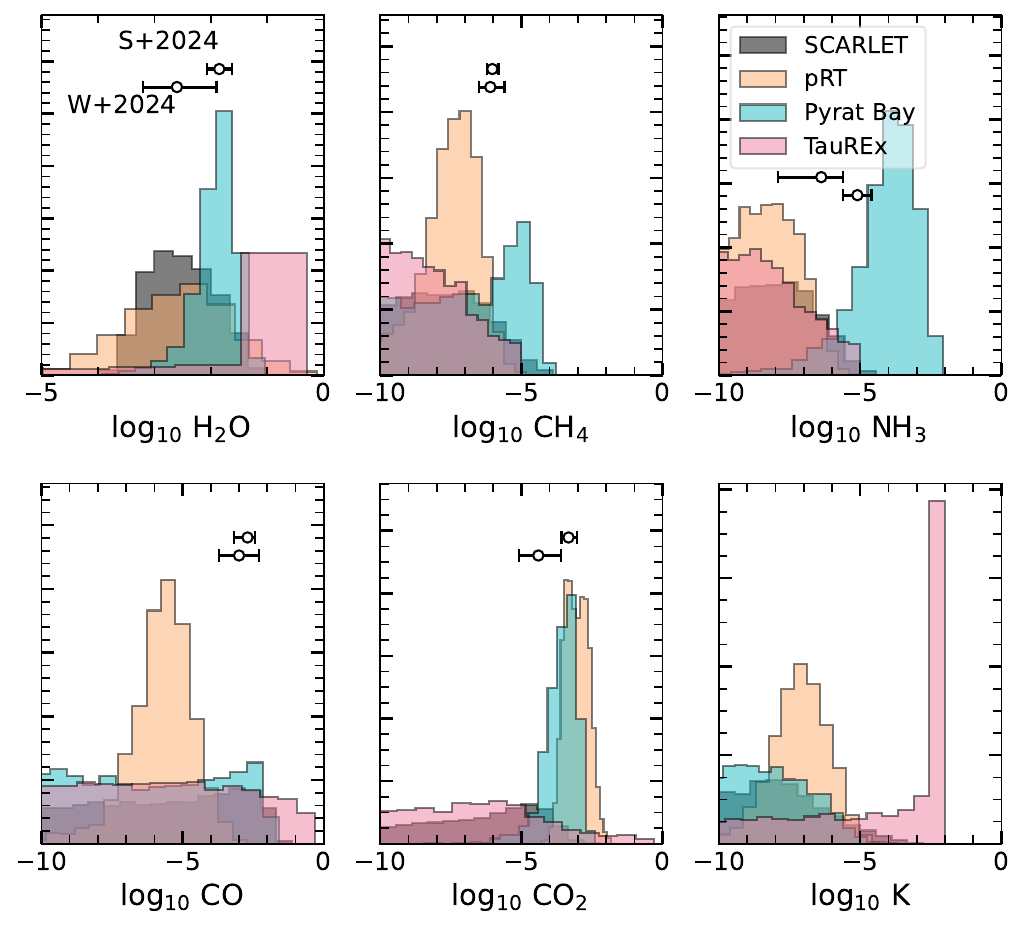}
  \caption{Same as Figure \ref{fig:posterior_molecules_compare} with posteriors from \texttt{TauREx} added.}
  \label{fig: posteriors_appendix_taurex_added}
\end{figure*}

Our \texttt{TauREx} retrieval provides a poor fit to the dataset, as the short-wavelength haze treatment in \texttt{TauREx} is strongly species-dependent. The model compensates by adjusting molecular abundances to fit the slope, and vice versa. Therefore, we do not interpret any results from this retrieval. Additionally, the lower resolution of the opacities used in \texttt{TauREx} may have contributed to the poor fit. Nevertheless, this exercise is valuable, as it highlights the complexity involved in interpreting the nature of the short-wavelength slope. The corner plots from \texttt{TauREx} are shown in Figure \ref{fig: corner taurex}, and the posteriors (see Figure \ref{fig: posteriors_appendix_taurex_added}) further illustrate the weak constraints it places on various molecular species.

\end{appendices}


\backmatter

\bmhead{Data and code availability}
The data and code used in our analysis will be available upon request.

\bmhead{Supplementary information}

The appendix section has all the additional information than in the main text.

\bmhead{Acknowledgements}

This project was undertaken with the financial support of the Canadian Space Agency. We thank L.W. for sharing the best-fit model from the published paper \citet{welbanks_high_2024}. C.P.-G. acknowledges support from the NSERC Vanier scholarship, and the Trottier Family Foundation. C.P.-G. also acknowledges support from the E. Margaret Burbidge Prize Postdoctoral Fellowship from the Brinson Foundation.
R.A.\ acknowledges the Swiss National Science Foundation (SNSF) support under the Post-Doc Mobility grant P500PT\_222212 and the support of the Institut Trottier de Recherche sur les Exoplanètes (iREx). D.J.\ is supported by NRC Canada and by an NSERC Discovery Grant. This work has been carried out within the framework of the NCCR PlanetS supported by the Swiss National Science Foundation under grants 51NF40$\_$182901 and 51NF40$\_$205606. S.P.\ acknowledges the financial support of the SNSF. JT acknowledges funding support by the TESS Guest Investigator Program G06165. This project has received funding from the European Research Council (ERC) under the European Union's Horizon 2020 research and innovation programme (project {\sc Spice Dune}, grant agreement No 947634).

\section*{Declarations}

Not applicable

\bibliography{sn-article}

\end{document}